\RequirePackage{fix-cm}
\documentclass[epjc3]{svjour3}  
\smartqed  
\RequirePackage{graphicx}
\journalname{Eur. Phys. J. E}
\begin{document}

\title{A stochastic model for dynamics of FtsZ filaments and the formation of $Z$-ring
}


\author{Arabind Swain\thanksref{e1,addr1,addr2,addr3}
        \and
        A. V. Anil Kumar\thanksref{e2,addr1,addr2}
        \and Sumedha\thanksref{e3,e4,addr1,addr2}
}

\thankstext{e1}{e-mail: arabind7swain@gmail.com}
\thankstext{e2}{e-mail: anil@niser.ac.in}
\thankstext{e3}{e-mail: sumedha@niser.ac.in}
\thankstext{e4}{corresponding author}


\institute{School of Physical Sciences, National Institute of Science Education and Research, Jatni - 752050, India\label{addr1}
           \and
    Homi Bhabha National Institute, Training School Complex, Anushakti Nagar, Mumbai 400094, India        \label{addr2}
           \and
           \emph{Present Address:} Department of Physics, Emory College of Arts and Sciences,Atlanta, Georgia 30322 USA \label{addr3}
}

\date{Received: date / Accepted: date}

\maketitle

\begin{abstract}

Understanding the mechanisms responsible for the formation and growth of FtsZ polymers and their subsequent
formation of the $Z$-ring is important for gaining insight into the cell division in prokaryotic cells. 
In this work, we present a minimal stochastic model that qualitatively reproduces {\it in vitro} observations of 
polymerization, formation of dynamic contractile ring that is stable for a long time and depolymerization 
shown by FtsZ polymer filaments. In this stochastic model, we explore different mechanisms for ring breaking and hydrolysis. 
In addition to hydrolysis, which is known to regulate the dynamics of other tubulin polymers like microtubules, we 
find that the presence of the ring allows for an additional mechanism 
for regulating the dynamics of FtsZ  polymers. Ring breaking dynamics in the presence of hydrolysis naturally induce 
rescue and catastrophe events in this model irrespective of the mechanism of hydrolysis. 

\end{abstract}

\section{Introduction}
Cell division is one of the most fundamental processes in living cells.  In prokaryotic cells, numerous proteins take part 
in assembling the machinery of cell division, called the divisome.
FtsZ(Filamenting temperature-sensitive mutant Z ) is the most important among them. This tubulin homologue polymerizes head 
to tail, forming dynamic protofilaments. These protofilaments together form the core of a structure called the Z-ring in 
the division plane of the cell\cite{nature1,J. Lutkenhaus,Eirikson1,zip,cross}. The Z-ring persists throughout the cell 
division. Recent studies on septal wall suggest that FtsZ monomers move around the ring via treadmilling which guides and 
regulates their growth, which in turn controls division \cite{tread1,tread2,Tread3,Tread4,vor3}).  This dynamic ring is a 
phenomenon that is intrinsic to FtsZ and the importance of this contractile ring in cell division is well established by now. 
Recently, there have been a lot of focus both {\it in vivo} \cite{Tread3,Tread4} and {\it in vitro} \cite{invitro18,afm3} 
to understand the dynamics of the ring. But the mechanism controlling the activity of the ring is still poorly understood. 

The monomers of FtsZ consist of two independent domains. The $N$-terminal domain with 
its parallel beta sheets connected to alpha helices, provides the binding site for GTP/GDP. The C- terminal region is 
essential for FtsZ to interact with other proteins like FtsA and ZipA\cite{ma1999} and also as a membrane tether for 
the Z-ring formation during cell division\cite{tetherC}. Above a critical concentration of GTP, FtsZ polymerizes 
cooperatively into filaments which are single stranded and have a head to tail orientation with polymerization at 
only one end \cite{Conc1,Conc2,conc3,conc4,conc5,conc8,conc9,conc10,conc11,conc12,polar}. The exact mechanism behind 
this cooperativity is not well understood. Studies have shown that FtsZ can polymerize into multi-stranded bundles, 
sheets and form chiral vortex on membranes \cite{vor3,conc12,vor1,vor2,vor4,mult1,mult2} depending on the assembly 
buffer, nucleotide and other proteins present.


Atomic Force Microscopy(AFM) of filaments adsorbed on Mica surface has shown the role played by lateral 
interactions and filament curvature in understanding the dynamic behavior of FtsZ filaments\cite{AFM,afm2}. The study of 
individual filaments under AFM showed that the filaments can form rings which are dynamic in nature. These rings may open 
up, lose material and close again until they open irreversibly and then the filaments depolymerize. It has been shown that 
unlike the case of tubulin, where nucleotide exchange is the rate limiting step for the reaction kinetics, in the case of 
FtsZ, it is the hydrolysis which controls the reaction kinetics. The ring formation of FtsZ is studied in the presence of GTP, 
guanosine-5’-[($\beta, \gamma$)-methyleno-triphosphate (GMPCPP) which reduces the hydrolysis as well as depolymerization 
rates and GTP with glycerol in the buffer which reduces the depolymerization rates and hence stabilizes the lifetime for 
irreversible opening\cite{afm3}.  They found that the rings formed in the presence of GMPCPP had broader length distribution, 
with a higher average length than the rings formed in the presence of GTP. A very recent study by 
Ramirez-Diaz et al \cite{invitro18}  has again focused on the ring formation {\it in vitro}. They made 
some key observations. They found that the fine tuning of hydrolysis is needed for 
the formation of stable rings. They also found that the treadmilling results from a directional growth of curved 
and polar filaments from the nucleation point at the membrane. The preferential addition of GTP subunit to the 
leading edge establishes a GTP-GDP gradient. 

FtsZ monomers are very similar to eucaryotic tubulin polymers like microtubules(MTs) and actin in their composition 
and functioning. MTs are semiflexible polymers that are a key component of the mitotic spindle\cite{mitoticspindle} 
in eukaryotic cells and this requires them to  be dynamic in nature. They exhibit a phenomena of dynamic instability, 
in which they switch from a phase of slow elongation to rapid shortening (catastrophe) and from rapid shortening to 
growth(rescue). A polymerizing MT grows until it suffers a catastrophe and starts to depolymerise. Similarly, a 
depolymerising MT undergoes rescue and starts polymerising again. Attempts using stochastic models have been 
successful in explaining this behaviour\cite{hill,bayley,vavylonis,leibler,antal}. Experimental results and 
theory together have now established that the hydrolysis of tubulin monomers is responsible for the dynamic 
instability of MTs, though the exact mechanism of hydrolysis is still debated \cite{howard1}. Similarly, actin 
filaments\cite{actin1,actin2} also exhibit non equilibrium phenomena known as treadmilling. In treadmilling, the 
new subunits get added at the growing end and old subunits leave the polymer from the other end. Mathematical models 
of MTs and actin have helped our understanding of the phenomena of dynamic instability and treadmilling. These models 
bridge scales and hence facilitate our understanding of complex biological phenomena in terms of elementary 
processes\cite{howard}. This helps in organizing the plethora of information one gets from the biochemistry study of 
these proteins.

The basic self-assembly mechanism underlying dynamic instability, assembly mediated by nucleotide 
phosphate activity, is omnipresent in biological systems. FtsZ falls in the same class of bio-polymers and hence it seems 
natural that a similar modelling approach capturing the dynamics will help our understanding of the FtsZ. Mechanical 
models\cite{pre09,pmc14,sm16,pre18}, based on torsion and curvature to explain experimental data have been attempted. 
But except for a few deterministic approaches to model the FtsZ ring, there have been almost no attemps at modelling 
FtsZ at the molecular level. Most of these deterministic models aimed to characterise theoretically the {\it in vivo} 
and {\it in vitro} observations of FtsZ assembly. These models involve a number of differential equations to be solved 
simultaneously making it computationally very costly. The eight equation model proposed by Chen {\it et al.}\cite{chen1,chen2} 
described the initial stages of FtsZ polymerization successfully, but fails to handle the whole process of FtsZ assembly. 
There are other models by Dow {\it et al.}\cite{dow}, Lan {\it et al.}\cite{lan} and Surovtsev {\it et al.}\cite{mod1} which 
employ few hundreds of differential equations making the computations very complex. Recently, kinetic models based on average 
charcteristics of different species and their concentrations have been proposed, where the number of differential equations 
need to be solved has reduced considerably to 17\cite{martinez1} or 10\cite{martinez2}. These models reduced the computational 
cost drastically and were able to predict the time taken to reach the steady state, the concentration of FtsZ in the Z-ring 
and average dimension of the filaments and bundles, which were in agreement with the experimental observations. However, the 
dependence of these on factors like rate of hydrolysis could not be obtained by these models. These deterministic models 
cannot capture the ring dynamics , which is stochastic in nature. There have been very few attempts to model the FtsZ ring 
formation stochastically\cite{gov1,gov2}. These studies concentrated on the clustering and ring formation of FtsZ polymers, by 
modeling the attachment and detachment of these polymers to the cell membrane. For a recent review on modelling FtsZ see \cite{review}.

MTs and actin form straight filaments while FtsZ form a ring of roughly one micro-meter diameter in its active state \cite{invitro18}. In this paper we propose a stochastic model for FtsZ with treadmilling, 
which is similar in spirit to known stochastic models of MTs and actin\cite{antal,howard1,actin1,actin2}. Our aim is to describe 
the FtsZ dynamics observed in the {\it in vitro} experiments.
The models on MTs and actins take into account three processes mainly: polymerisation, depolymerisation and hydrolysis. Since the {\it in vitro} experiments reveal that the ring formation in FtsZ polymers is dynamic in nature, we introduce additional processes to account for ring formation and breaking in the model. We consider two mechanisms for 
ring opening : a) random breaking of the ring, i.e., the ring opens randomly at any interface and b) non-random breaking of 
the ring, i.e., rupture can occur only at the interface with atleast one GDP bound FtsZ monomer. We also consider both 
possible mechanisms of hydrolysis of tubulin monomers: namely the vectorial and stochastic hydrolysis \cite{howard1}. 
In vectorial hydrolysis, an unhydrolyzed monomer gets hydrolyzed only if the neighboring monomer is already hydrolyzed. 
This is a highly cooperative mechanism and there exists a sharp boundary between the hydrolyzed and unhydrolyzed parts of 
the FtsZ filaments. In the stochastic hydrolysis, GTP-FtsZ subunit can hydrolyze in a stochastic manner, irrespective of 
the position of the subunit in the protofilament. The rate of hydrolysis in this case would be proportional to the amount 
of unhydrolyzed monomers. We find that the hydrolysis of the monomers is essential for the dynamics of FtsZ polymer. But 
interestingly the rescue and catastrophe events are more crucially regulated by the process responsible for opening and 
closing of the Z-ring.  Hence we consider four models in this paper based on the nature of hydrolysis and the mechanism 
of ring breaking: vectorial hydrolysis with random ring breaking, vectorial hydrolysis with non-random ring breaking, 
stochastic hydrolysis with random ring breaking and stochastic hydrolysis with non-random ring breaking. The stability 
of ring improves with randomness and stochastic hydrolysis with random ring breaking gives the most stable rings. We find 
that in the case of random ring breaking, we need to fine tune hydrolysis rate to get a stable ring like structure. On the 
other hand, the dynamics in the case on non-random breaking is insensitive to the hydrolysis rate. Also the experimentally 
observed ring length distribution matches with the ones obtained by vectorial hydrolysis with random ring breaking. Hence 
we conclude that the non-random ring breaking mechanism can be ruled out. This agrees with earlier suggestions \cite{afm3} 
and also with the suggestions that ring breaks due to tension created on the ring due to deformation of the 
membrane \cite{forcepinch,force11}.


The plan of the paper is as follows: In Section \ref{model} we describe our model. In Section \ref{vechyd}  
we consider the vectorial hydrolysis with random and non-random breaking of the ring and find that random ring 
breaking gives rise to stable dynamic ring. In section \ref{ringbreak} we study the effect of changing the ring 
breaking rate on the dynamics and conclude that with the random ring breaking one gets a filament that remains stable 
for a long time and then contracts. In section \ref{sh} we look at the stochastic hydrolysis with random and non-random 
ring breaking and compare it with the results of vectorial hydrolysis. We discuss our 
results in Section \ref{concl}.

\section{\label{model}Model}
\begin{center}
\begin{figure}[htp]
\includegraphics[width=0.8\textwidth]{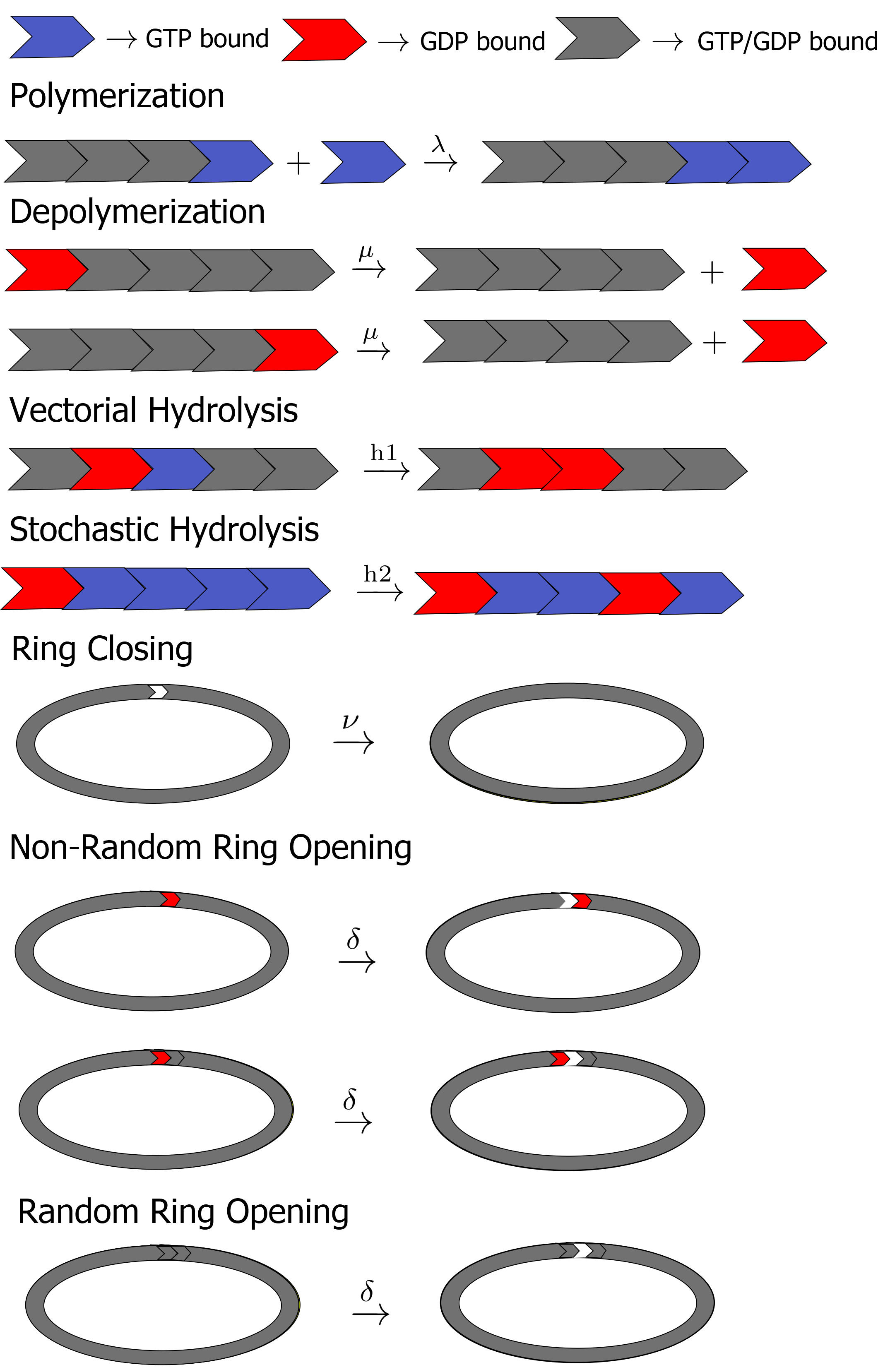}
\caption{Schematics for the different reactions taking place in the system.}
\label{fig:scheme}
\end{figure}
\end{center}

FtsZ filament is known to exist both in the form of an open chain and as a closed ring. In the open form it has a single active end(the arrow head) which is in contact with a reservoir with GTP-bound monomers. Based on recent experimental evidence \cite{vor3,invitro18,polar}, we assume the filament to be directional with polymerization only possible at its head end. In our model, in the presence of a GTP-bound monomer at its polymerizing end, a GTP-bound monomer gets added to the polymer at a rate $\lambda$. Whenever a GDP-bound monomer is at either ends of an open filament, the polymer can lose a GDP-bound monomer from the ends with a rate $\mu$. Both these processes are independent. This introduces treadmilling in the model. 
It has been experimentally observed that depolymerization is faster at the ends as compared to the center\cite{afm3}. Hence, we have assumed depolymerization only at 
the ends.  Also, in general the polymerization rate is higher than the 
depolymerization rate\cite{vor3}. 

When polymerized FtsZ in a solution of GTP is exposed to an excess of GDP it depolymerizes quickly\cite{gdp1}. GTP-bound monomers have been found to constitute $80\%$ of filaments under in vitro conditions\cite{conc3,gdp2}.  Though GDP-bound monomers have been found to polymerize in vitro but as compared to polymerization of GTP-bound monomers the equilibrium constant for this process has been found to be significantly lower\cite{conc8}. These results suggest that the GDP-bound polymerization is unlikely to be viable and thus irrelevant in explaining FtsZ dynamics. Hence , in our model we assume that there is no polymerization of GDP-bound monomers.

Our minimal stochastic model incorporates the process of polymerization at the preferred end, 
depolymerization from both ends, hydrolysis, ring breaking and closing. 
In Figure \ref{fig:scheme}, we present the schematic of all the processes we include in our model. 
 
The cases for both vectorial and stochastic hydrolysis have been considered individually. In the case of vectorial hydrolysis, the interface grows by converting the GTP-bound monomer associated with a GDP-GTP interface to a GDP-bound monomer with a rate $h_{1}$. Propensity of a reaction in stochastic processes gives us the likelihood of a particular reaction happening in a unit time. Reactions with higher propensities are more likely. Thus the propensity for this reaction is
\begin{equation}
h{vec}=h_{1}\times (Number~of~GDP-GTP~interfaces)
\end{equation}
In the case of stochastic hydrolysis, a GTP-bound monomer can get randomly converted into a GDP-bound monomer with 
a rate $h_{2}$. The propensity for the reaction is 
\begin{equation}
h_{rand}=h_{2} \times (Number~of~GTP~bound~monomers)
\end{equation} 
The hydrolysis reactions can take place both in the open as well as the ring forming filaments.     

Following the scheme used in earlier works\cite{mod1,mod3} to model the natural curvature of FtsZ we assume that the filament has an optimum length($N_0$) at which it has the maximum probability of closing and forming a ring. Hence, given a length $N$, the polymer has a non-zero probability to form a ring. We assume that the rate of this process depends on $N$ and is 
given by a Gaussian distribution that peaks at $N=N_0$: 
\begin{equation}
\nu=\frac{C_0}{\sigma\sqrt{2\pi}}e^{-\frac{(N-N_0)^2}{2 \sigma^2}}
\end{equation}

FtsZ is known to polymerize into single stranded protofilaments 
{\it{in-vitro}} \cite{Conc2}. Since treadmilling occurs at the level of a single protofilament, we have not included lateral interaction in our present model. 

Ring opening was assumed to have two possibilities. Firstly, the ring opens at any monomer-monomer interface chosen randomly. 
If we take $\delta$ to be the rate of random breaking, then $U_2=\delta N$ is the propensity of this reaction. Secondly, given that the GDP associated bonds are much weaker than GTP associated bonds, we also consider the case where the ring breaking is only possible at an interface containing atleast one GDP-bound monomer. So, for a given ring with $N_1$ GDP bound interfaces, the propensity of breaking is $U_3=\delta N_1$.

The dynamics hence can be described by a set of coupled chemical master equations for the probability of open 
and closed polymer. Due to the presence of ring breaking, it is not possible to solve the equations even in 
the simpler case of vectorial hydrolysis. Hence we use Gillespie algorithm\cite{Gillepsie1,Gillepsie2} to 
solve the chemical master equations numerically. Gillespie algorithm offers an elegant way to speed up simulations 
by doing away with the many rejected trials of the  traditional Monte-Carlo moves. While, traditional Monte Carlo methods 
check at each step if each reaction takes place, Gillespie algorithm draws directly the next reaction and time elapsed until 
that next reaction. The advantage of Gillespie algorithm is that it generates an ensemble of trajectories with the correct 
statistics. It has been very successful in simulating many chemical and biological reactions \cite{Gillepsie2}. In fact, Gillespie
alogorithm was employed before in modelling the clustering and ring formation of FtsZ polymers\cite{gov1,gov2}.

\section{Results and Discussion}

\subsection{\label{vechyd}Vectorial Hydrolysis}

In this section we study the dynamics of FtsZ rings when the hydrolysis is assumed to be only at the interface, i.e 
the vectorial hydrolysis.  Let $P_r(N,t)$ represents the probability of having a closed ring polymer of length $N$ at 
time $t$ and $P_o(N,t)$ represents the probability of having an open chain polymer of length $N$ at time $t$. We 
define $p_1(t)$ as the probability of having a GTP-bound monomer at the active end in the open ring configuration. 
Thus the probability of active end having a GDP bound monomer is $1-p_1(t)$. Let $p_2(t)$ be the probability of 
finding a GDP bound monomer at the non-polymerizing end. The probability of observing a GDP monomer at either end 
is $p_3(t)=p_2(t)+1-p_1(t)$ ,  As discussed in Section \ref{model},  
$\nu =\frac{C_0}{\sigma\sqrt{2\pi}}e^{\frac{-(N-N_0)^2}{2 \sigma^2}}$ is the rate for an open polymer of 
length $N$ to close and form a ring. Then for random breaking the system can be described by the following set of equations: 
\begin{equation}
\frac{\partial P_r(N,t)}{\partial t}=\nu P_o(N,t) -\delta NP_r(N,t)
\end{equation}
\begin{eqnarray}
\frac{\partial P_o(N,t)}{\partial t}&=& -p_1(t)\lambda\frac{\partial P_o(N,t)}{\partial N}+p_3(t)\mu \frac{\partial P_o(N,t)}{\partial N}\\ \nonumber & &-\nu P_o(N,t)+\delta NP_r(N,t)
\end{eqnarray}

In the beginning there are only the processes of polymerization, depolymerization from the back end and hydrolysis.
Once FtsZ polymer attains sufficient length, it can form a closed ring. In the ring form the polymer doesn't have a 
polymerizing and a non-polymerizing end. In this state there are only two reactions going on :hydrolysis and ring opening. 
Thus polymerization rate does not 
have a direct effect on the dynamics of a closed ring. The ring opening being a random process, it is not possible 
to tell what will the ends of the polymer be when it opens up. Hence, once the ring opens up, it can undergo 
polymerization or depolymerization depending on the type of monomers present at the ends.  This randomness in the opening of the ring makes it possible for the polymer to undergo dynamic 
catastrophe and rescue events. This makes the ring length and the ratio of GDP/GTP bound monomers in the ring fluctuating parameters, which in turn makes the calculation of $p_1(t)$ and $p_3(t)$ impossible. Hence, it is not possible to 
solve the chemical master equations  exactly. One can solve the equations exactly only in the trivial case where 
hydrolysis is much lower than the polymerization rate and hence the probability $p_1(t)$ and $p_3(t)$ can be taken 
to be $1$ and $0$ respectively. In this case one gets a gaussian distribution for the  ring length distribution. 
We hence simulate the system using Gillespie algorithm. We will compare the 
mechanisms of ring opening i.e random breaking and non-random breaking by comparing the properties of 
experimentally possible observables like ring length and ring life time.  We fix depolymerisation rate 
to be 1 (as we can always fix one of the rates to be one) and take $N_0=120$ and $\sigma=15$. $C_0$ is chosen 
such that $\frac{C_0}{\sigma \sqrt{2 \pi}}$ is $10$. We will now study the effect of polymerisation,hydrolysis 
and ring breaking rate on the dynamics of the ring. 



\begin{figure}
\includegraphics[width=0.5\textwidth]{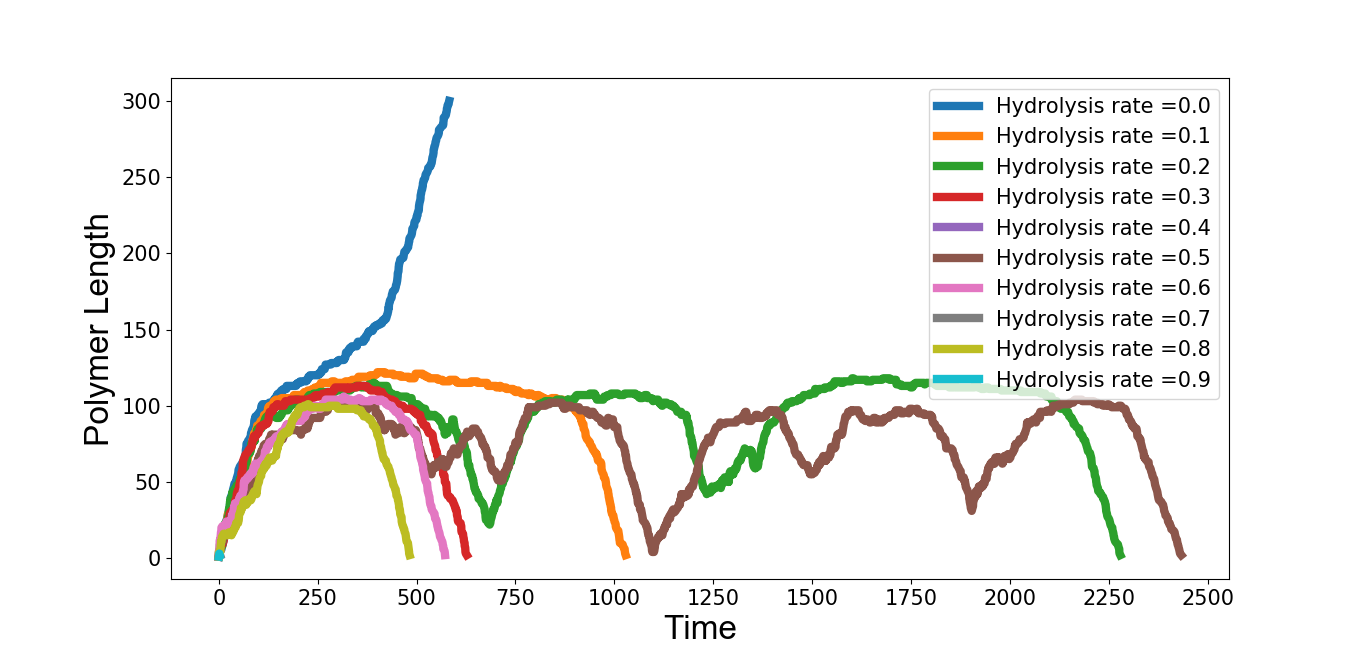}
\includegraphics[width=0.5\textwidth]{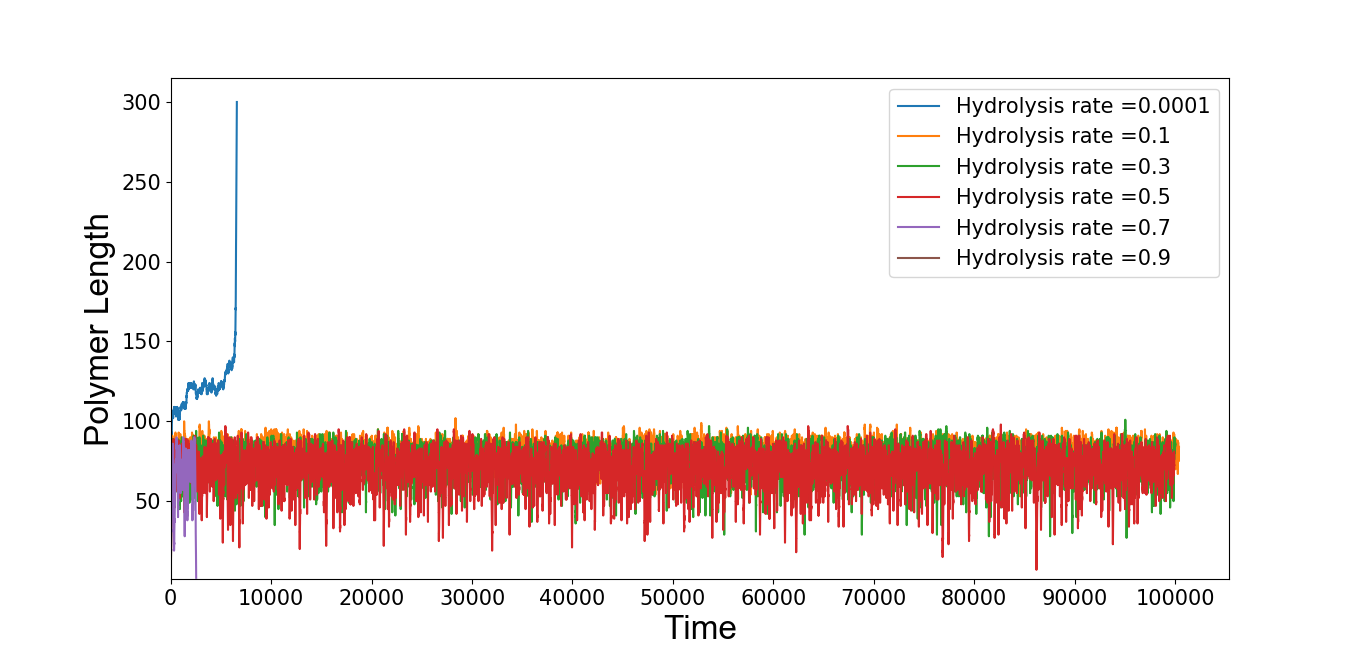}
\includegraphics[width=0.5\textwidth]{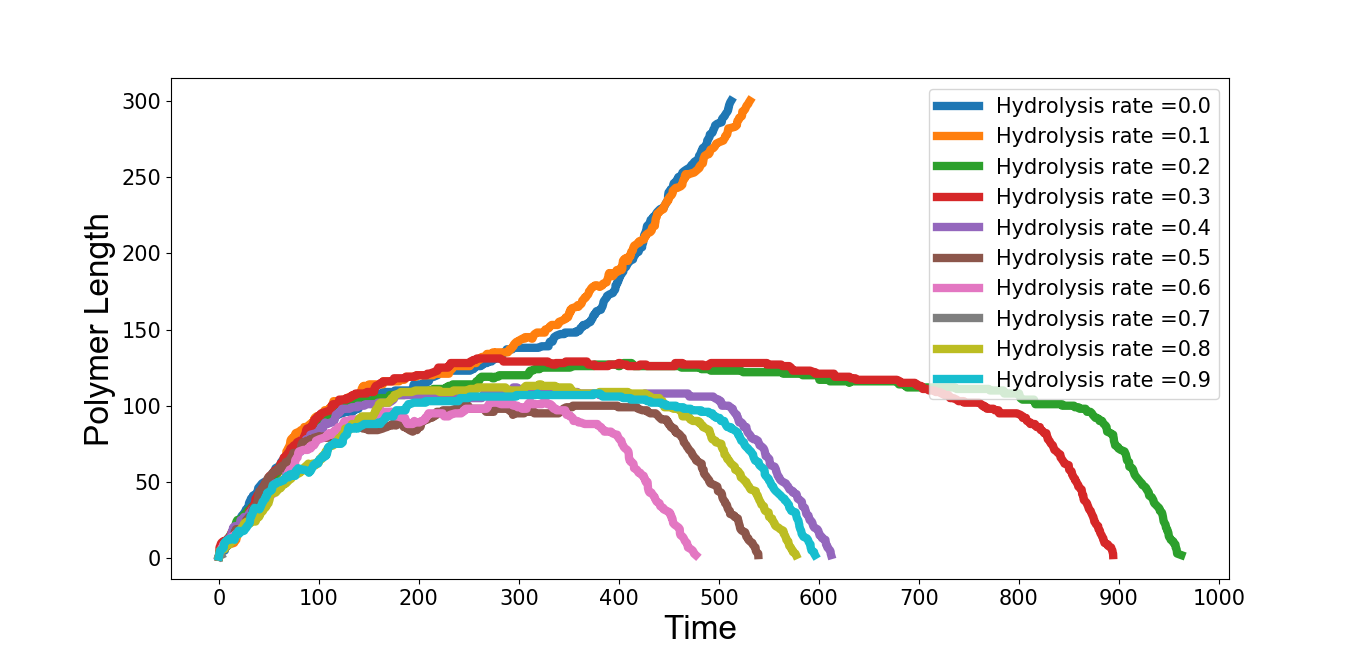}
\includegraphics[width=0.5\textwidth]{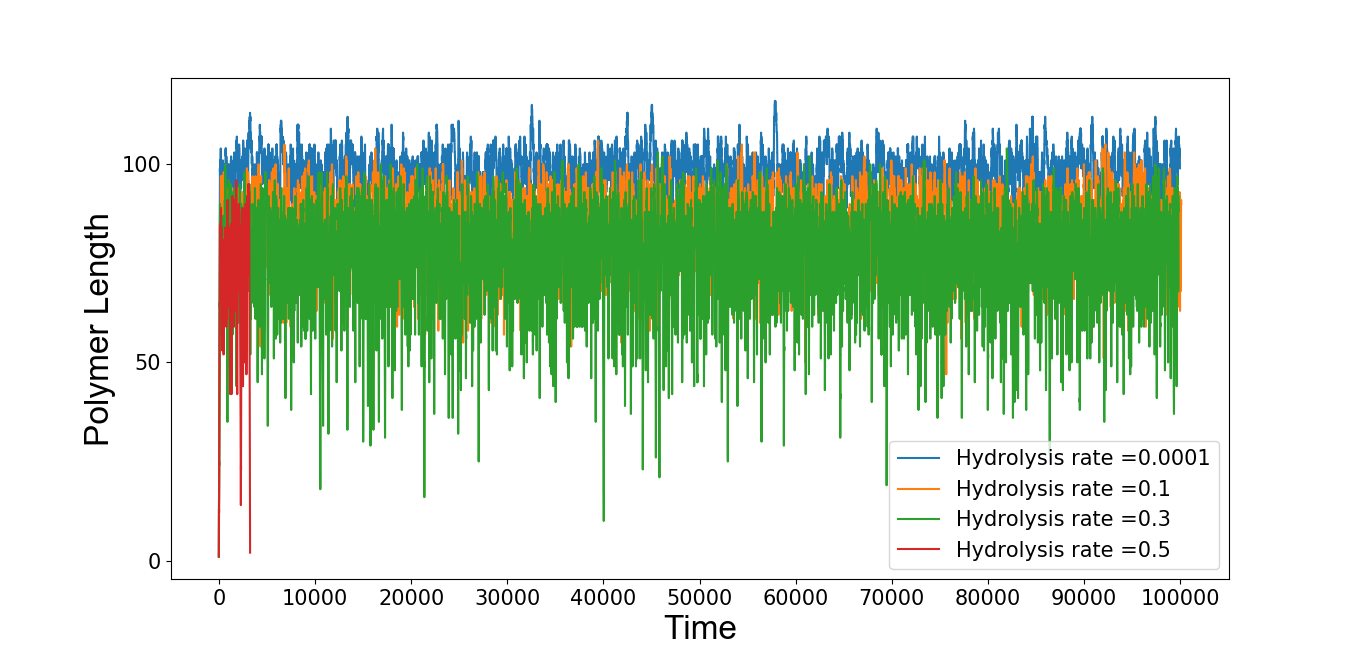},
\includegraphics[width=0.5\textwidth]{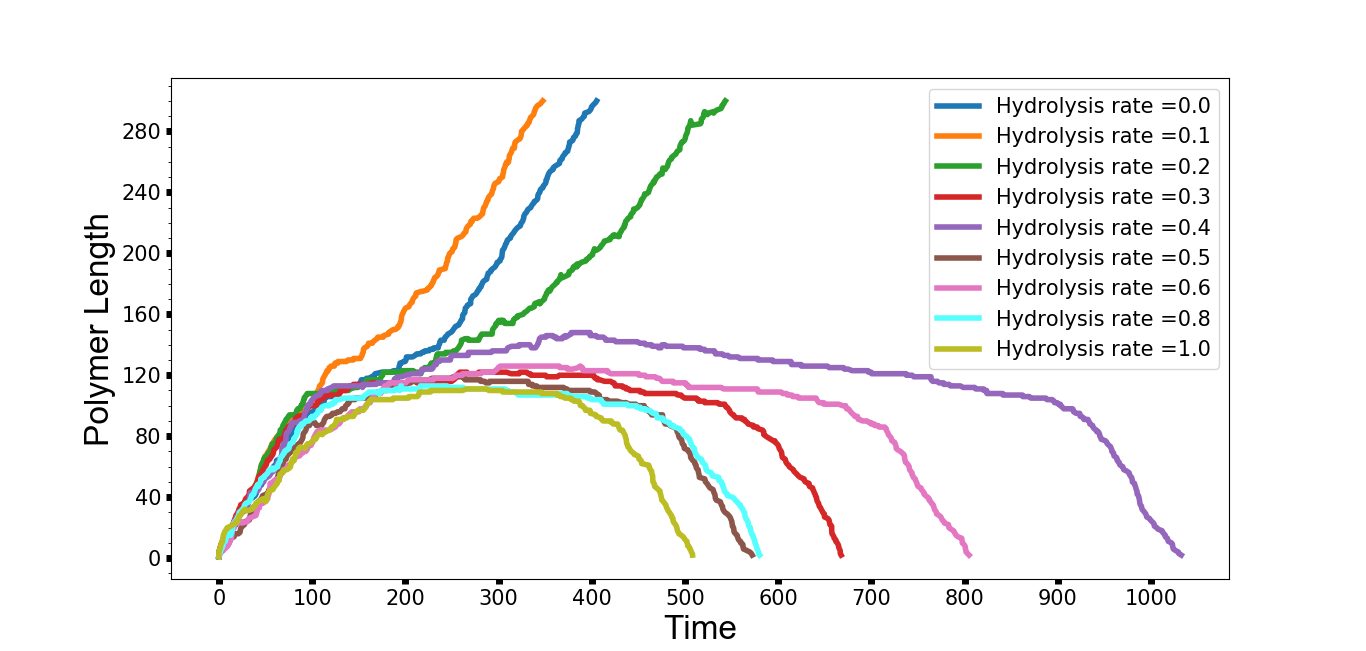}
\includegraphics[width=0.5\textwidth]{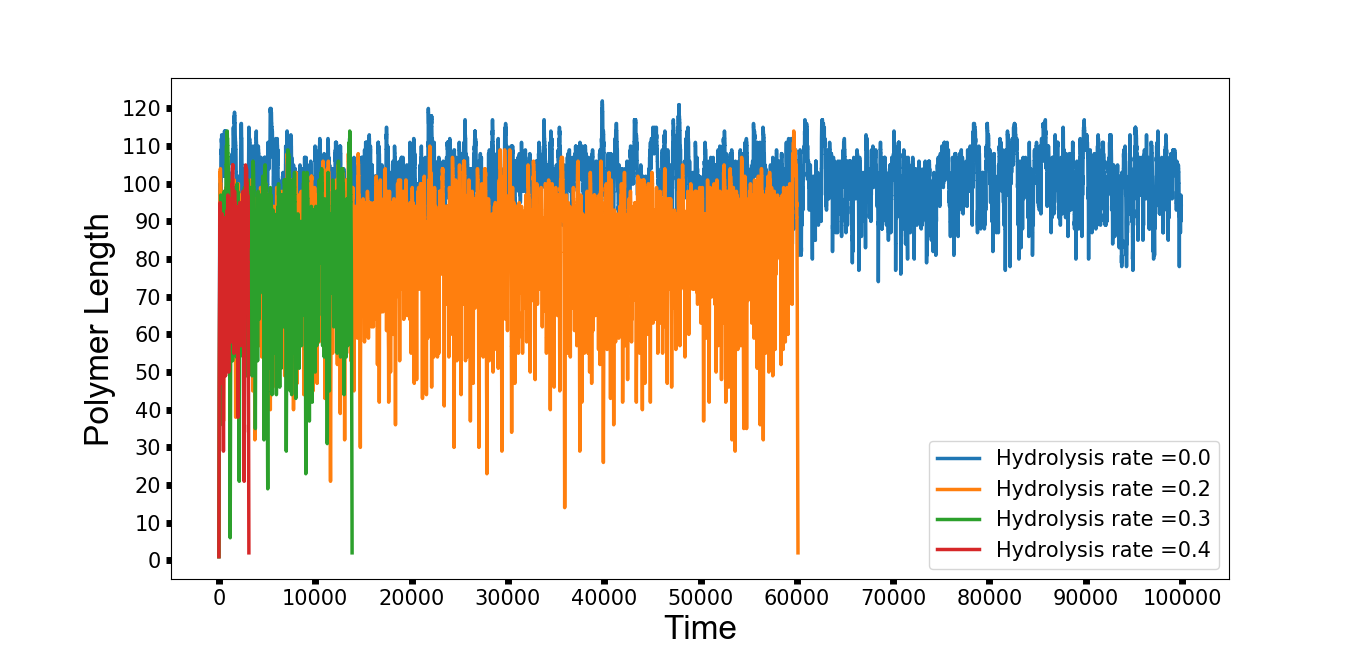}
\caption{Trajectories for Random(left) and Non-Random Breaking(right) for polymerization rate 2.0, 4.0 and 8.0 respectively(top to bottom). The depolymerisation rate($\mu$) is taken to be $1$ and ring breaking rate ($\delta$) is $0.008$: In the random breaking(left) case, the polymer initially undergoes a growth phase, in which the length of the polymer grows linearly with time. Depending on the hydrolysis rate($h_1$), it continues to grow or forms a dynamic ring. The ring formed stays dynamic for a long time. In the non-random case(right), a ring is formed for non zero hydrolysis rate and its dynamics is not sensitive to change in hydrolysis rate.}
\label{trajectories}
\end{figure}

In Figure \ref{trajectories}(left) we plot the length of FtsZ polymer as a function of time from a typical run for different hydrolysis rates for random breaking of FtsZ rings. We fix the random breaking rate to be $\delta=0.008$. The figure suggests that there are three distinct regions. Initially, there is a growth phase, where the polymer chain is formed from FtsZ monomers. In this phase, the length of the polymer grows linearly with time and is insensitive to hydrolysis rate, $h_1$. After this initial growth phase, the polymer makes a transition to a phase where the length gets stabilised for a period. This is the phase where FtsZ ring exists. It is clear from the trajectories that FtsZ ring is dynamic as the length fluctuates with time. The third region is where the polymer either grows an open chain or depolymerise completely. 

\begin{figure}
\includegraphics[height=3.25 cm]{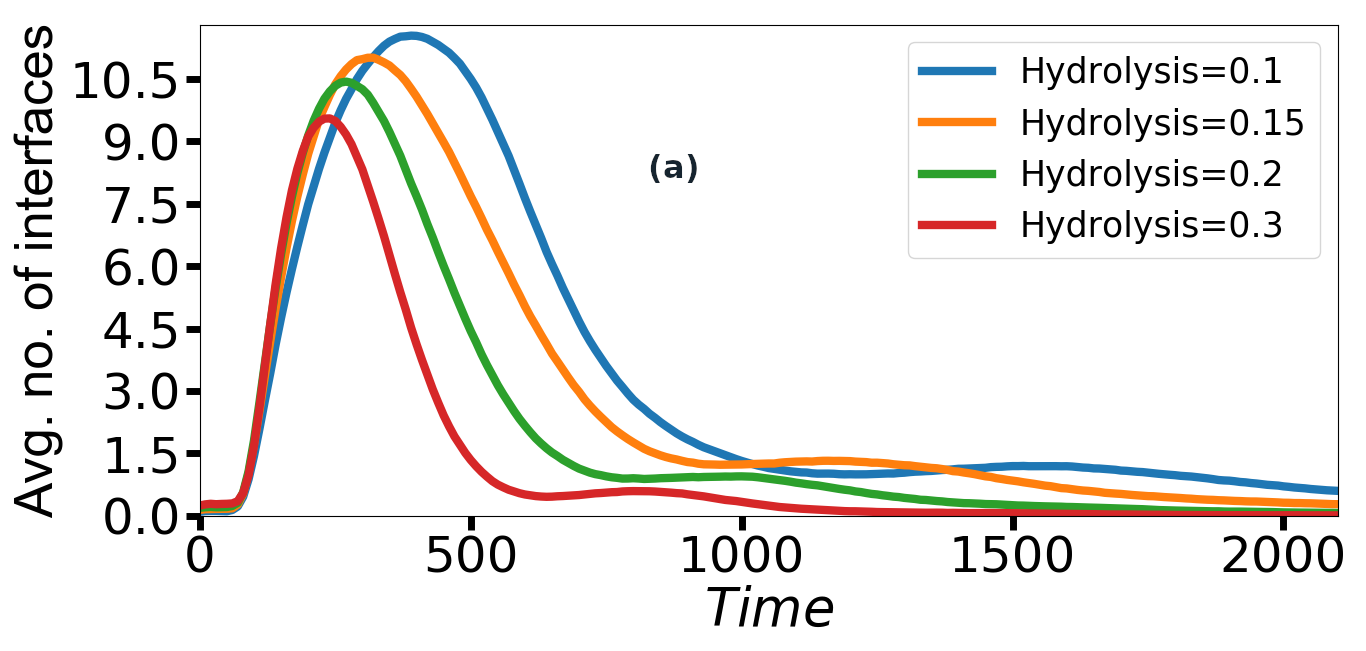}
\includegraphics[height=3.25 cm]{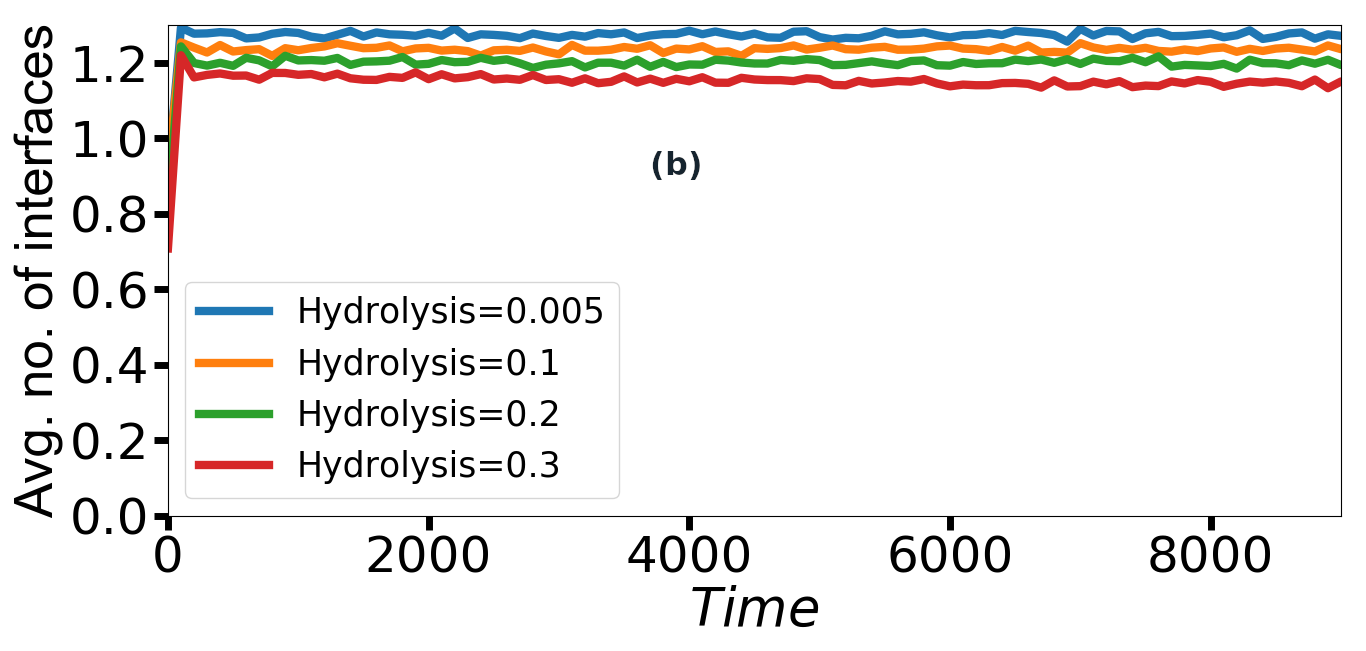}
\caption{Average number of D-T interfaces for random(a) and 
non-random(b) breaking for polymerisation rate $\lambda=4$. In the random ring breaking(a), the ring develops many more interfaces while active.}
\label{interface}
\end{figure}

When we consider random breaking, ring can break at any of the interface. 
When $h_1=0$, hydrolysis does not occur as there are no GDP bound monomers in the polymer. So when the ring opens up, it 
cannot depolymerise. Hence, after a short period of dynamic ring formation and breaking, the chain grows as a linear chain. 
When the hydrolysis rate is not zero, but small there is a finite probability to encounter a GDP bound monomer at one 
of the ends whenever the ring opens up. In this case, depolymerisation occurs before the chain start polymerising again 
resulting in a larger lifetime for the dynamic ring. As the hydrolysis 
rate increases, the probabilty of finding GDP bound monomer increases. In the case where GDP bound monomers are present 
at the polymerizing end, all of them need to get depolymerized to expose a GTP bound monomer for the polymer to  grow 
again. We find that this results in catastrophe and rescue events in which the polymer may lose as many as 
20-30 monomers before recovering again. Additionally, the opening and closing dynamics ensures that we have many interfaces 
in this case(see Fig. \ref{interface}(a)). Finally, when the polymer has no more GTP bound monomers and has only GDP bound 
monomers there are no more rescue events and the polymer depolymerizes completely. We find that as hydrolysis rate increases, initially the ring life 
time increases. However, because of interplay between polymerization, depolymerization, ring closing and opening, ring life time peaks 
at a finite value of $h_1$ and beyond that stability of ring keeps on decreasing. 

In Figure \ref{trajectories}(right), we plot the length of FtsZ polymers in the case of nonrandom 
breaking; i.e. the ring opens only at the interface formed by atleast one GDP bound 
monomer, for different hydrolysis rates. In the zero hydrolysis case once the ring gets 
formed it will never open since GDP bound interfaces are absent in the ring. At small values of hydrolysis rate, GTP will 
get hydrolyzed after a long time and at some later time 
it will open. Since GDP bound monomers are present at least at one of the ends, the chain is exposed for depolymerisation 
once it opens up. Hence, chances of depolymerization is bigger in this case in contrast to the random breaking case. This 
results in the large fluctations seen in the trajectories compared to the case of random breaking. The number of interfaces 
for vectorial hydrolysis remains the same, i.e, 1 throughout the time evolution(recall that for random breaking the number 
of interfaces increase with time). We have plotted the average number of interfaces as a function of time for different 
hydrolysis rates in Fig. \ref{interface}(b).


From Figure \ref{trajectories}, it is clear that the ring dynamics and its stability are sensitive to the hydrolysis rate
in the case of random breaking of the ring and insensitive to hydrolysis rate in the case of non-random breaking.
These observations can be made more quantitative in terms of experimentally measurable quantities like 
ring lifetime and length distributions. We obtain these quantities by averaging over $10^5$ independent trajectories. 

\subsubsection{Ring length distribution}
\begin{figure}
\centering
\includegraphics[width=0.48\textwidth,height=0.15\textheight]{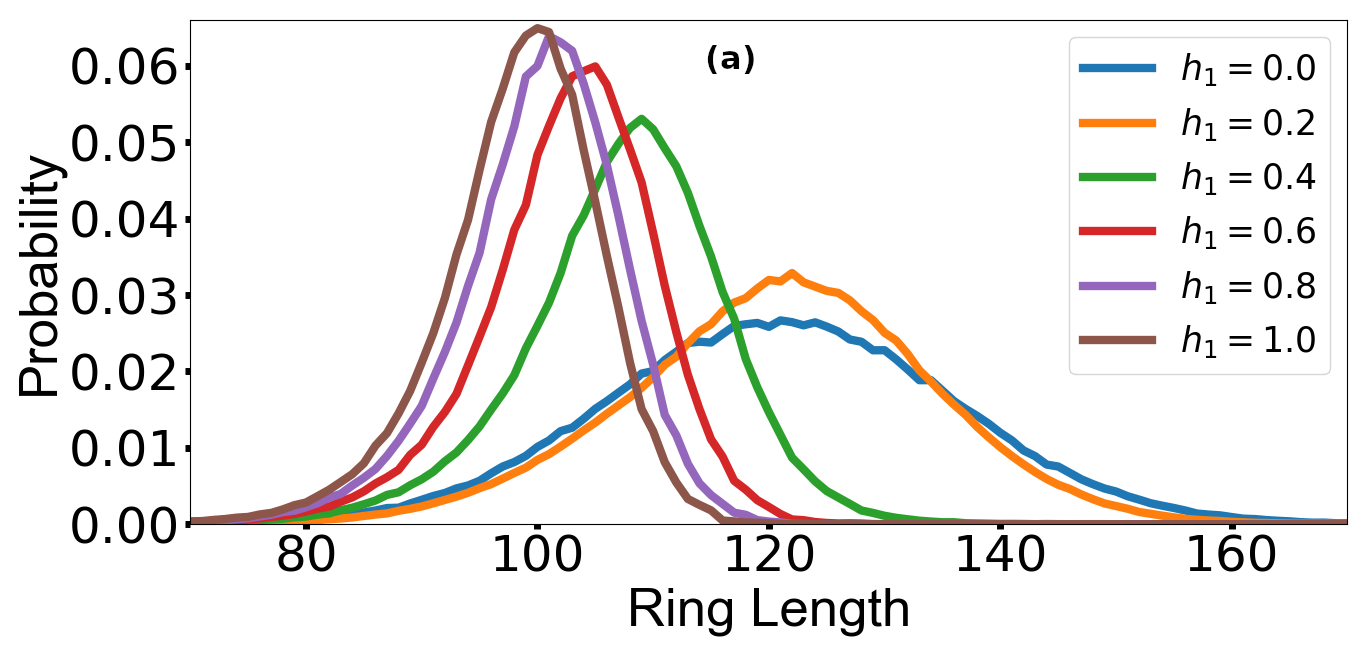}
\includegraphics[width=0.48\textwidth,height=0.15\textheight]{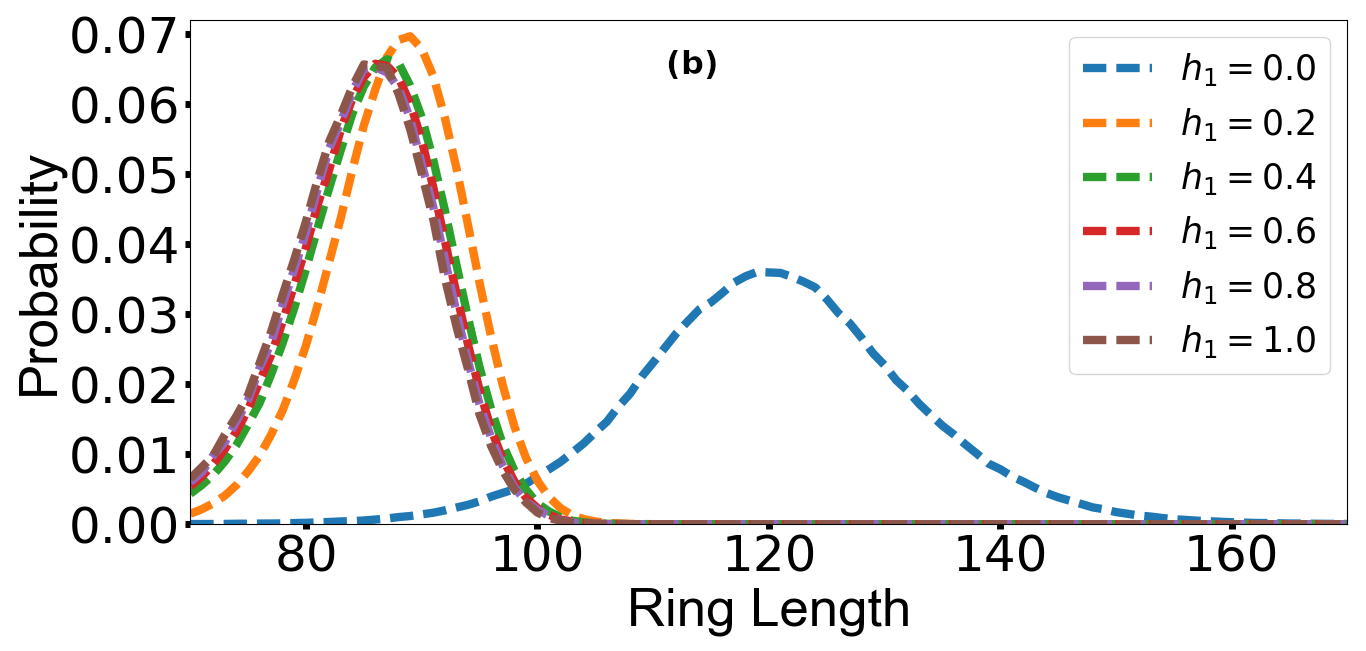}
\includegraphics[width=0.48\textwidth,height=0.15\textheight]{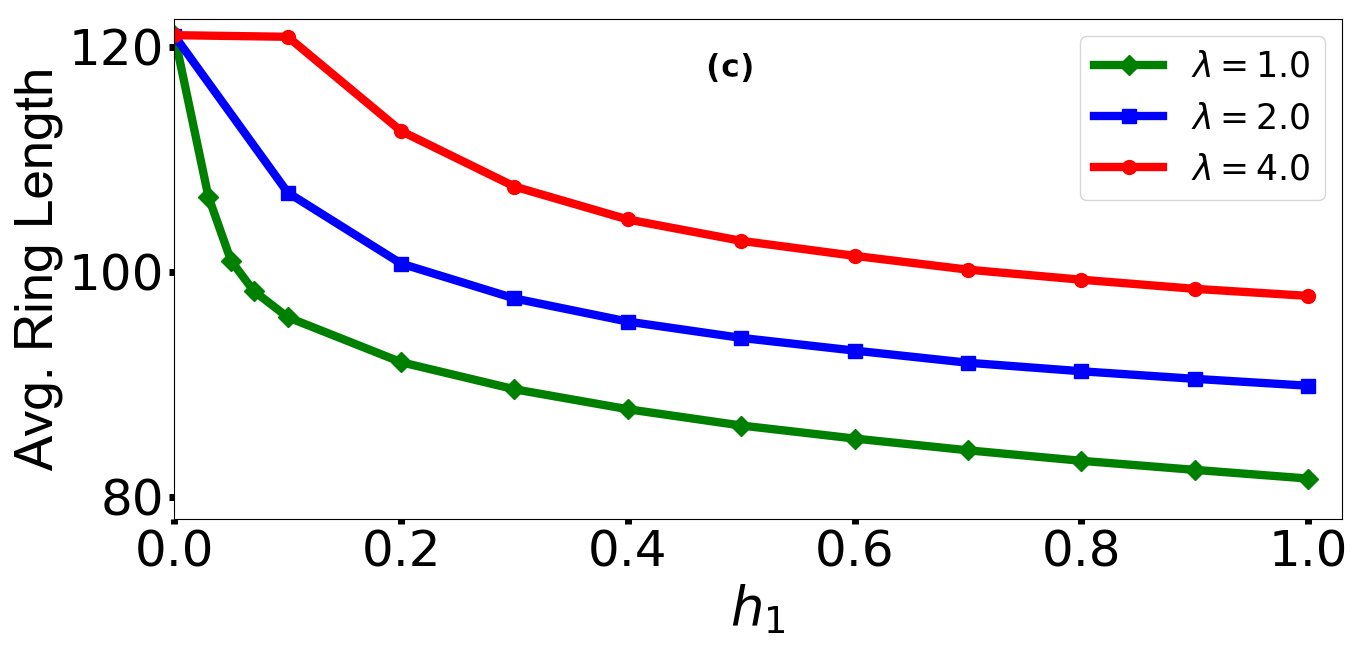}
\includegraphics[width=0.48\textwidth,height=0.15\textheight]{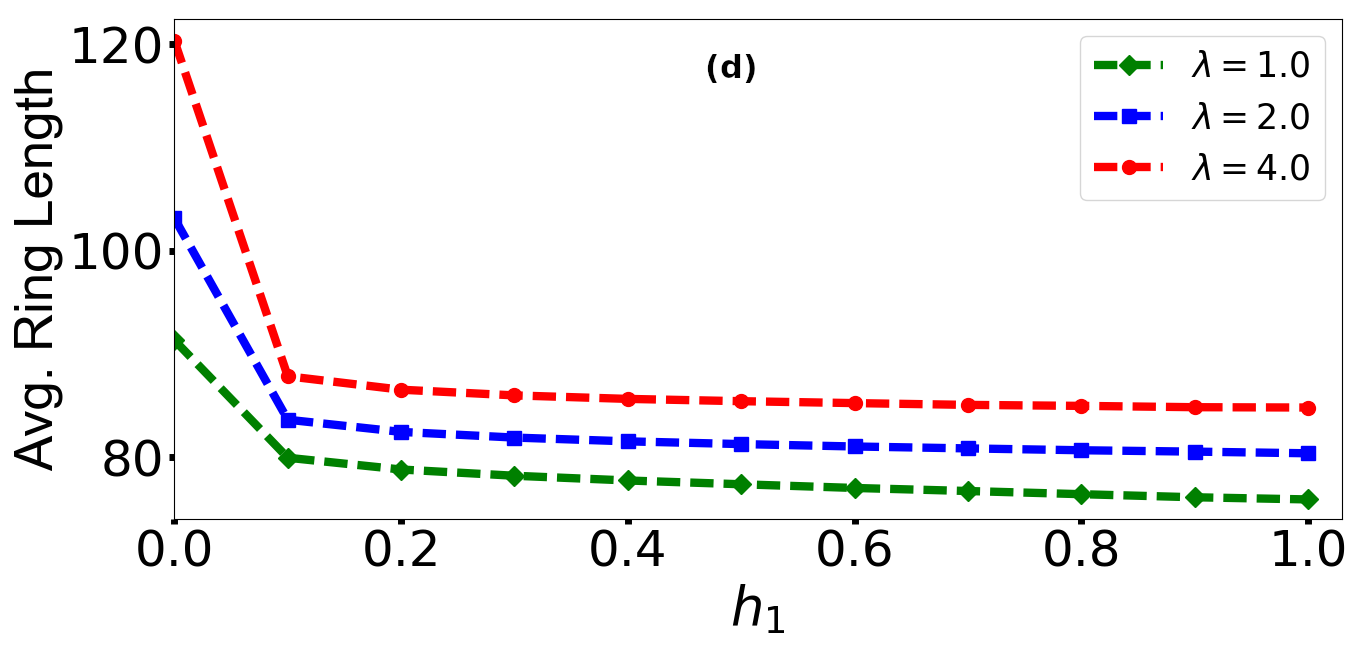}
\caption{a) and b): Plot of distribution of ring length for polymerization rate 4.0  for random and non-random ring breaking respectively. c) and d): Plot of average length for different  polymerization rates, for random and non-random breaking respectively}
\label{length}
\end{figure}

In Fig \ref{length}(a) and \ref{length}(b) we plot distributions of ring size in both random breaking and non-random breaking respectively for different hydrolysis rates. In the case of random breaking the distribution gets narrower with increasing hydrolysis rate. In ref. \cite{afm3} authors studied the ring length distribution of FtsZ polymers in-vitro. They found that rings formed in the presence of slowly hydrolyzing analogue, GMPCPP, resulted in broader distribution , which was shifted to the right. Our plots in Fig. \ref{length}(a) are consistent with this observation. The ring length distribution gets thinner with increasing hydrolysis rate and the peak shifts to the left, resulting in decreasing average length of the ring. In contrast, in the case of the non-random ring breaking, except for zero hydrolysis rate, the ring length distribution remains unchanged, suggesting that the ring length is insensitive to the changes in hydrolysis rate. In the case of zero hydrolysis rate, the ring never opens up once it forms as there are no GTP-GDP interfaces, and the distribution is just proportional to the ring closing rate $\nu$.  

In fig \ref{length}(c) and \ref{length}(d) we plot the average length of FtsZ rings in the two cases. As the hydrolysis rates increases, the average length of the polymer decreases gradually in the case of random breaking. In the case of non-random breaking, the average length remains same except for the case of zero hydrolysis rate. This suggests that the observation that the dynamics of FtsZ polymer is insensitive to the hydrolysis rate in the case of non-random breaking, made 
based on figure \ref{trajectories}(b) is indeed true. Near zero hydrolysis rate, the average length of ring, depends only on the 
polymerisation rate as the number of GDP-bound monomers are very less and depolymerisation does not take place often.
So the average length of the ring increases as the rate of polymerization increases.

\subsubsection{Time for irreversible ring opening}
\begin{figure}
\includegraphics[height=3.55 cm]{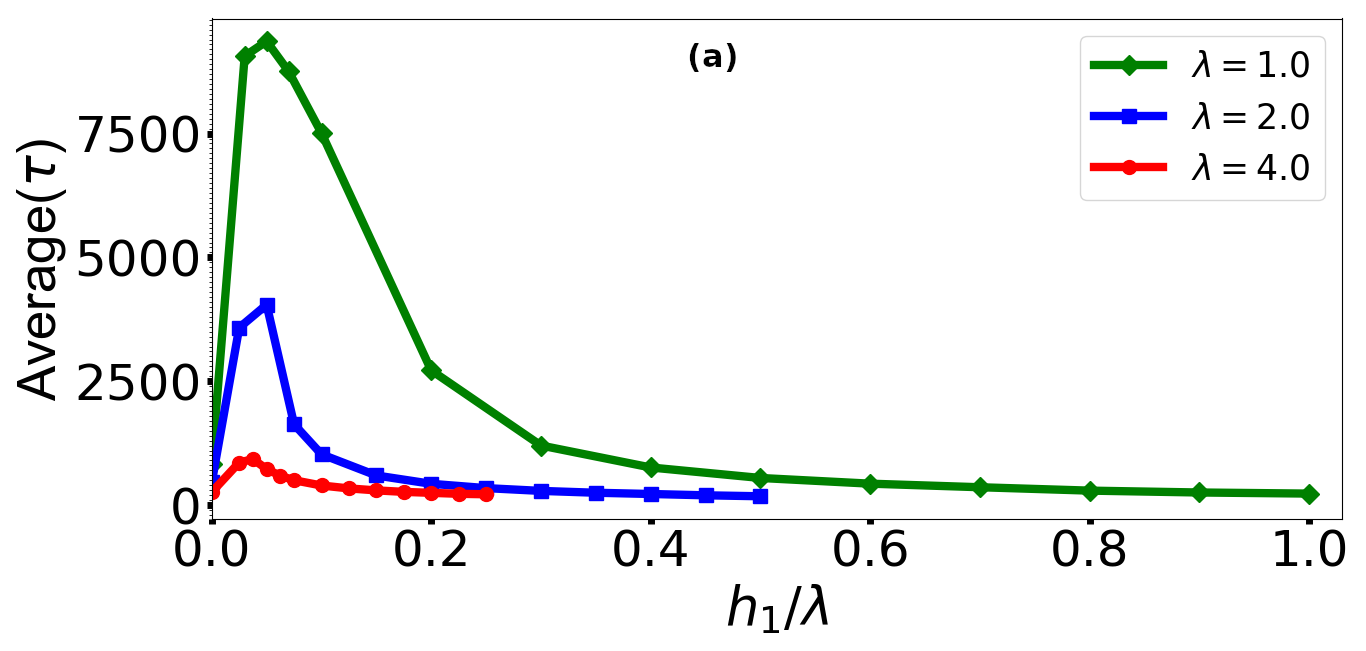}
\includegraphics[height=3.5 cm]{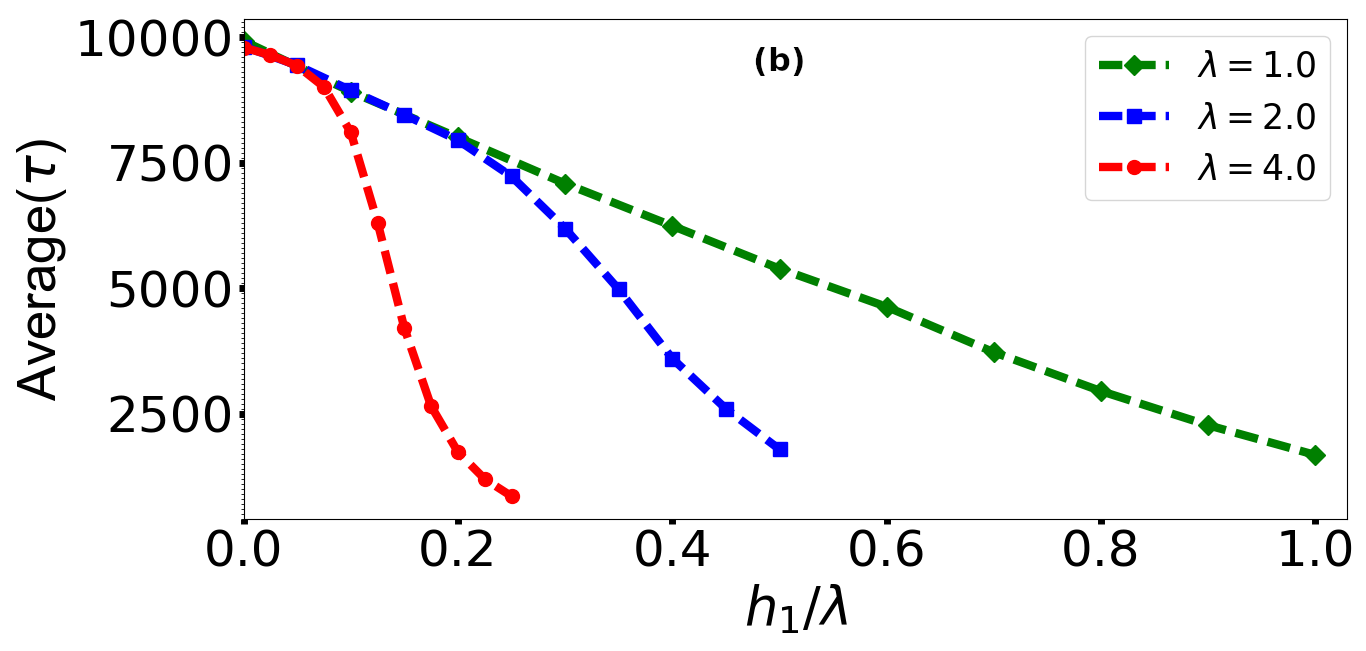}
\caption{Average lifetime($\tau$) for irreversible opening for random(a) and 
non-random(b) breaking. Here $\tau$ is the time for irreversible opening. For random ring breaking, $\tau$ is sensitive to the hydrolysis rate and peaks at a fixed value of $h_1/\lambda$,}
\label{lifetime}
\end{figure}

It is known that the dynamic ring stage of FtsZ polymer is very stable. The filament undergoes many catastrophe and 
rescue event during this time, before eventually decaying. Hence, one of the most important observable is the total 
time($\tau$) for which the ring exists before it eventually escapes the ring formation and grows as a linear chain or decays 
completely. For each trajectory the time for irreversible ring opening is calculated as $\tau = T_{final} - T_{init}$, where
$T_{final}$ is the time at which the ring opens up finally and afterwhich the polymer grows as a linear chain 
or decay completely and $T_{init}$ is the time at which the polymer forms a ring for the first time after the initial
growth phase.
This time is averaged over many independent trajectories. The time for this irreversible ring opening,
$\tau$ has been plotted in Figures \ref{lifetime}(a) and \ref{lifetime}(b) for random and non-random ring breaking respectively. For random breaking it peaks at a  finite value of hydrolysis rate. This is because at very low values of hydrolysis the polymer escapes into growth phase and at large value of hydrolysis, the chances of ring closing again is small. Hence 
there is a narrow window of hydrolysis rate where the dynamic ring is most stable. This 
is consistent with the very recent experiments of Ramirez-Diaz et al \cite{invitro18} where they found that fine tuning of hydrolysis was needed to get stable dynamic rings. The stability of individual filaments which was observed to be very sensitive to the hydrolysis rate may be a major contributing factor to ring stability observed in experiments. Moreover, 
we observe that the region of maximum stability occurs roughly at a fixed ratio of hydrolysis to polymerization rates. It has been observed that in the presence of a GTP regeneration system, the dynamic ring can continue for a long time\cite{gdp1}. This along with the fact that FtsZ remains highly 
conserved across species\cite{ftszcon}, can be used to argue that FtsZ sits on the edge of stability in its naturally occurring form. This is clearly demonstrated in the plots for random ring opening in which, across all polymerization rates the region of maximum 
stability happens at a  fixed ratio of hydrolysis and polymerization rates. In 
contrast to the case of random ring opening, for non-random breaking the ring is not 
really dynamic and hence the lifetime goes down monotonically with the hydrolysis rate. 


\subsubsection{Ring lifetime distribution}

\begin{figure}
\includegraphics[width=0.5\textwidth,height=0.15\textheight]{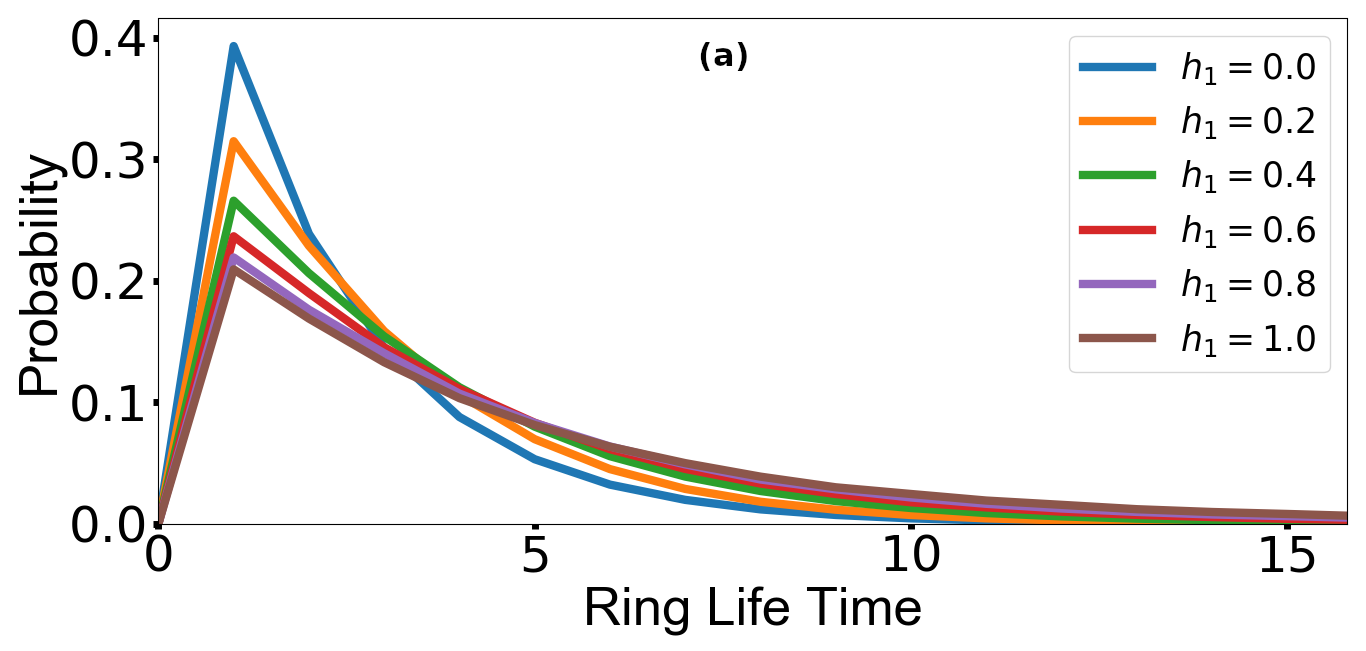}
\includegraphics[width=0.5\textwidth,height=0.15\textheight]{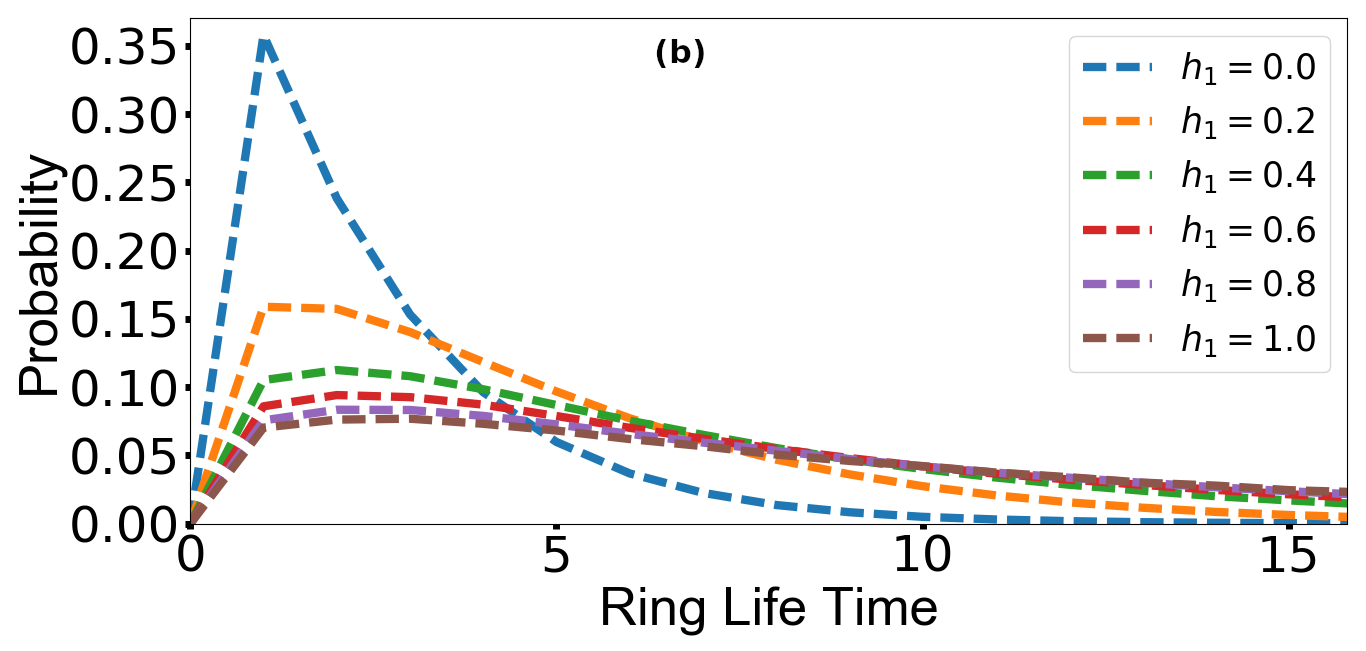}
\includegraphics[width=0.5\textwidth,height=0.15\textheight]{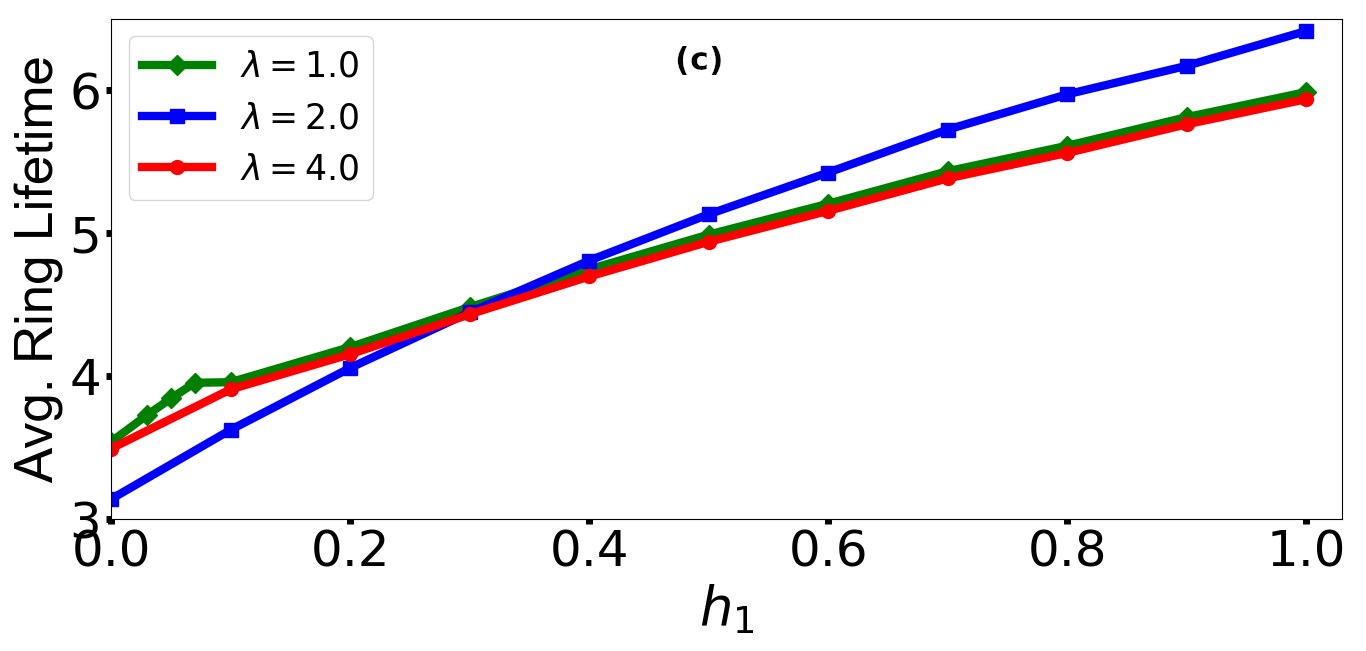}
\includegraphics[width=0.5\textwidth,height=0.15\textheight]{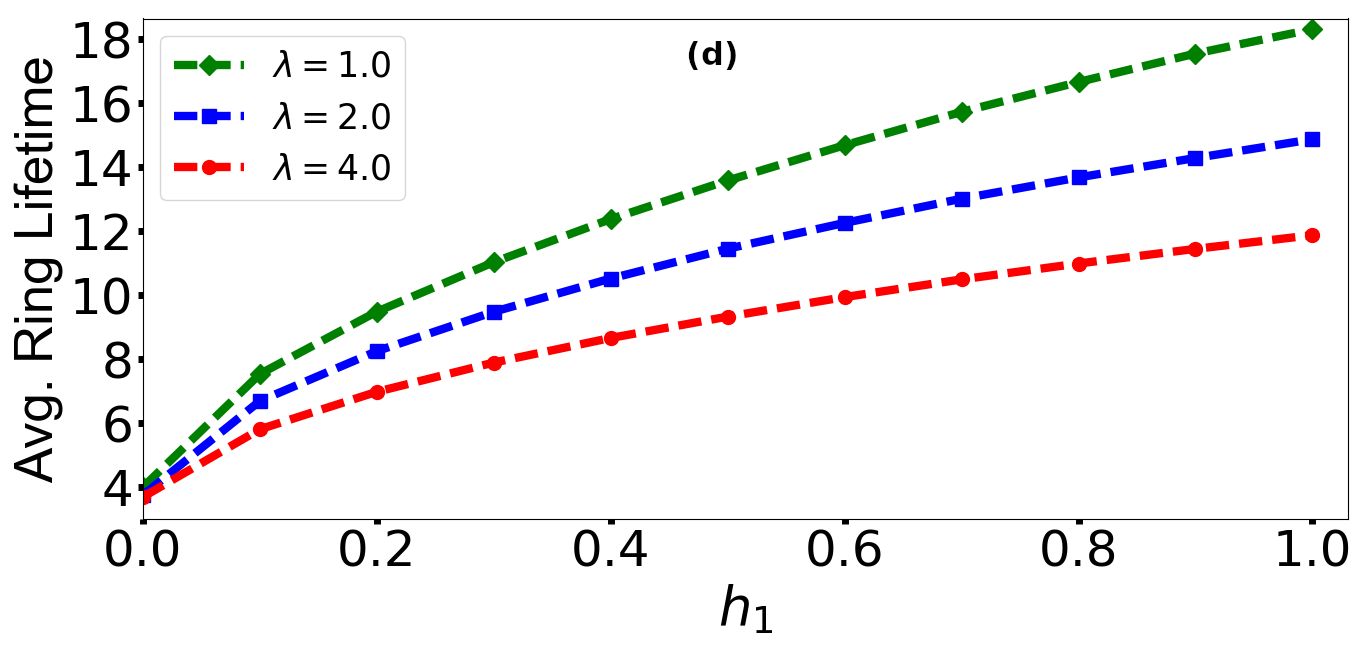}
\caption{a) and b): Plot of distribution of ring lifetime for polymerization rate 4.0 for random and non random ring breaking respectively. c) and d): Plot of average ring lifetime as a function of hydrolysis for random and non-random breaking respectively for different polymerisation rate. }
\label{ringlifetime}
\end{figure}

We can also look at the distribution of time elapsed between two openings. We expect that there would be many 
small events where the ring opens with GTP at the growing end and close back immediately. Figure \ref{ringlifetime}(a) and \ref{ringlifetime}(b)
show these distributions for the case of random and non-random ring breaking. The distributions become flatter and 
flatter as the hydrolysis rate increases for both cases; however the effect is more pronounced in the case of non-random breaking. 
Polymerization rate does not affect the lifetime of single openings since the polymerisation
does not take place when the ring is closed. 
Such a trend can be observed in the case of random ring opening. But in case of 
non-random opening, increase in polymerization rate leads to an decrease in average ring lifetime for the same hydrolysis rate.  Hence, we conclude that in the case of non-random opening polymerization rate affects the ring opening dynamics(see Fig. \ref{ringlifetime} (c) and \ref{ringlifetime}(d)).    

\subsubsection{Ring stability for random breaking}
\begin{figure}
\begin{center}
\includegraphics[width=0.8\textwidth]{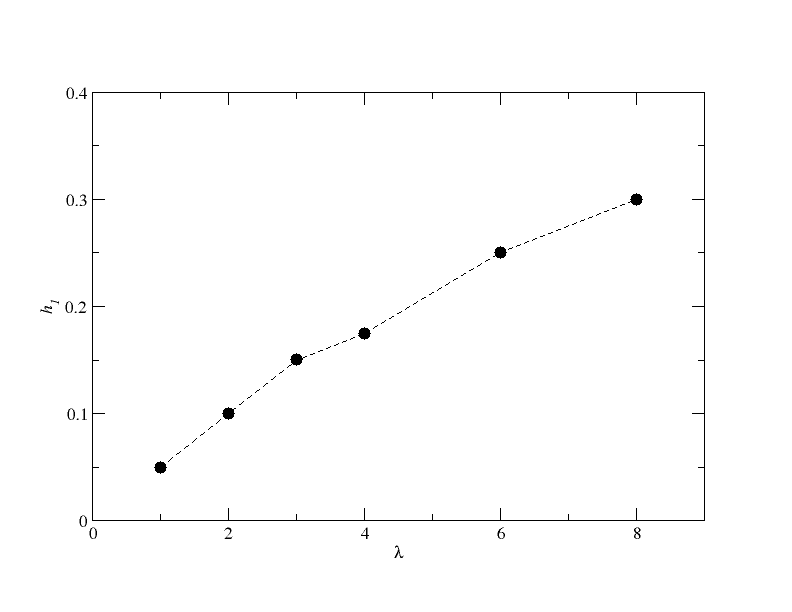}
\caption{The system can exist in 2 states. In one state the polymer keeps growing uncontrollably and forms a ring for a short time. In another state the polymer forms the ring before depolymerizing completely. Between these two extremes there are states in which the ring remains stable for a long time going through multiple events of ring closing, ring opening, polmerizations and depolymerizations. The plot shows the critical hydrolysis rate ($h_1$) at which the system has stable ring for the longest time before getting completely depolymerized as a function of polymerization rate ($\lambda$).}
\label{Phase}
\end{center}
\end{figure}

 From our observations on the average time for irreversible opening, it is clear that most stable ring is formed by fine 
 tuning the hydrolysis rate in the case of vectorial hydrolysis with random breaking. We find that for random breaking, the average life time($\tau$) does not change monotonically with the hydrolysis rate($h_1$) for a given polymerisation rate($\lambda$). In Fig. \ref{Phase}, we have plotted the value of $h_1$ for which $\tau$ is maximum as a function of polymerisation rate($\lambda$). We find that $h_1$ value for which $\tau$ becomes maximum increases with increasing $\lambda$.
 

\subsection{\label{ringbreak}Force generation and the effect of ring breaking rate on the dynamics}
\begin{figure}
\includegraphics[width=0.5\textwidth]{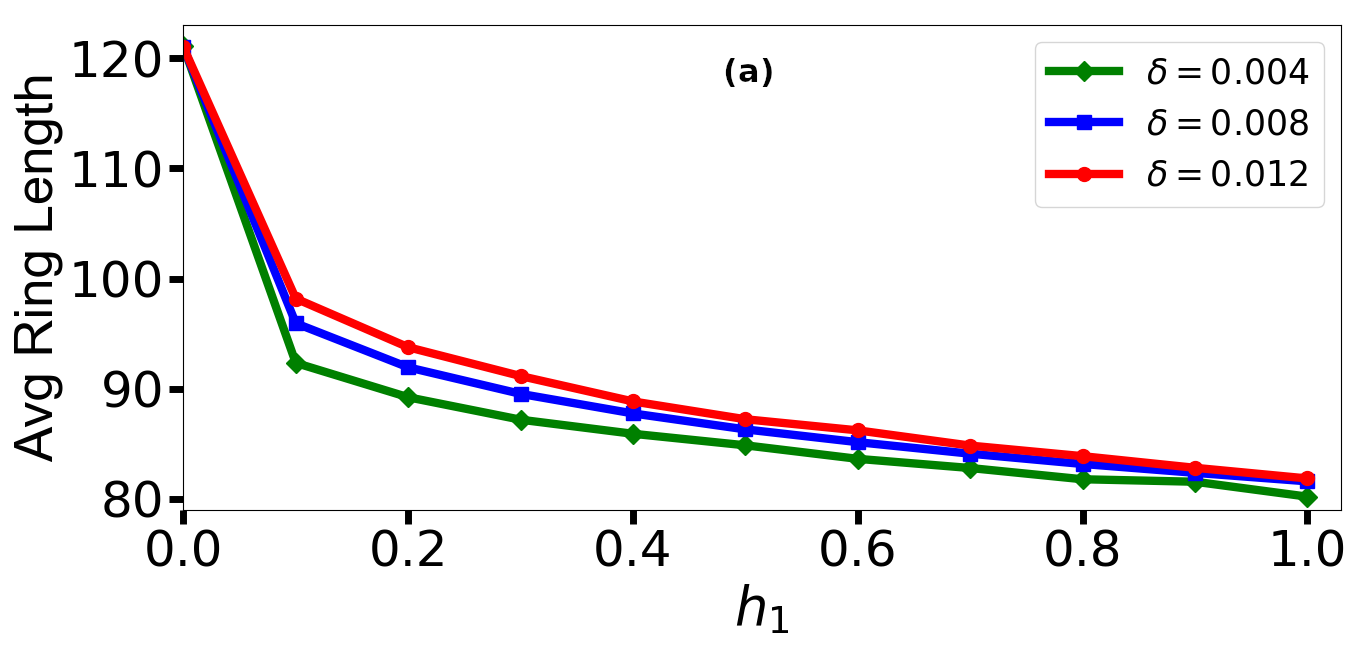}
\includegraphics[width=0.5\textwidth]{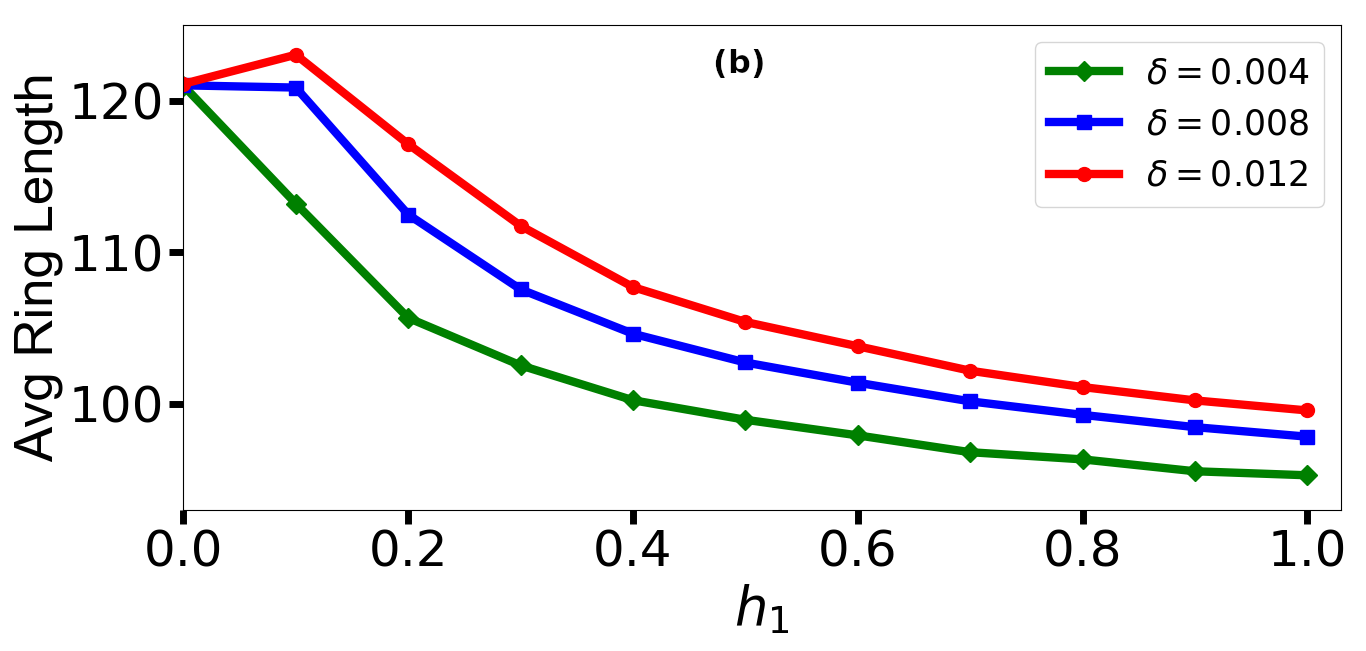}
\caption{Average lengths for different ring breaking rates $\delta$. The average ring length are plotted as a function of hydrolysis rates. Data for polymerization rate 1.0 and for polymerization rate 4.0 are plotted in (a) and (b) respectively. All the plots are for random breaking and the different breaking rates $\delta$ are represented by different colors.}
\label{for1}
\end{figure}

\begin{figure}
\includegraphics[width=0.5\textwidth]{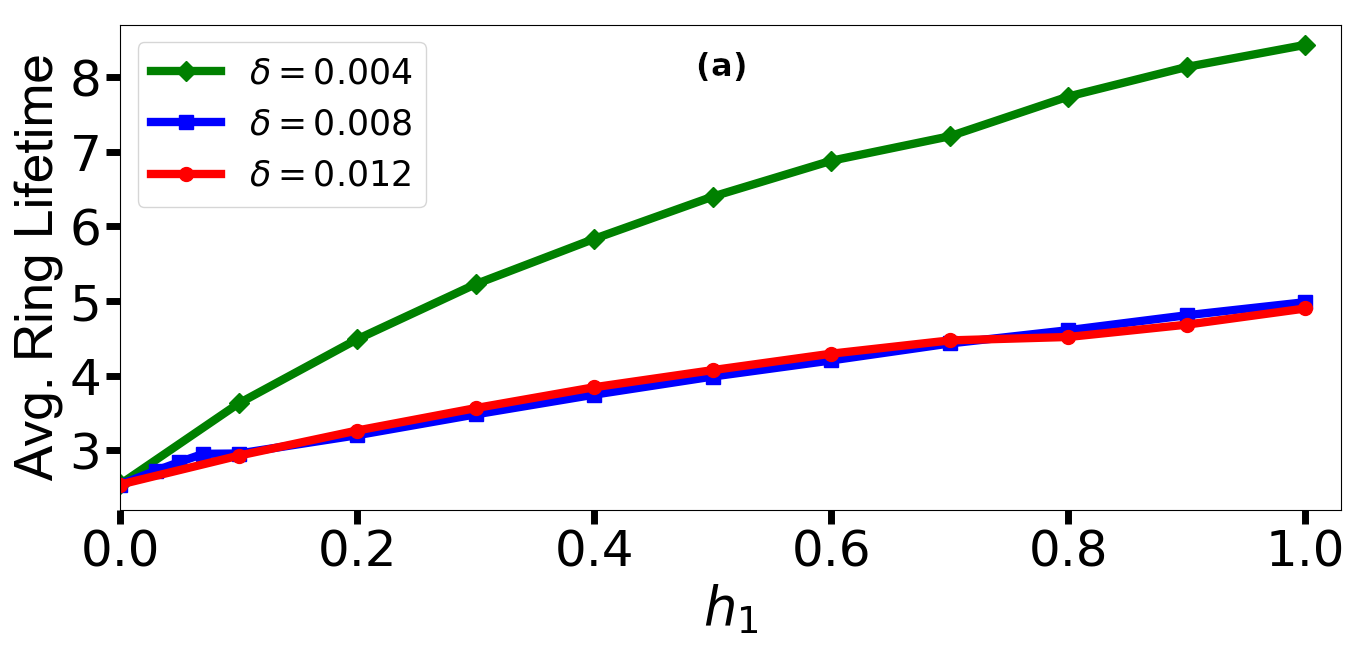}
\includegraphics[width=0.5\textwidth]{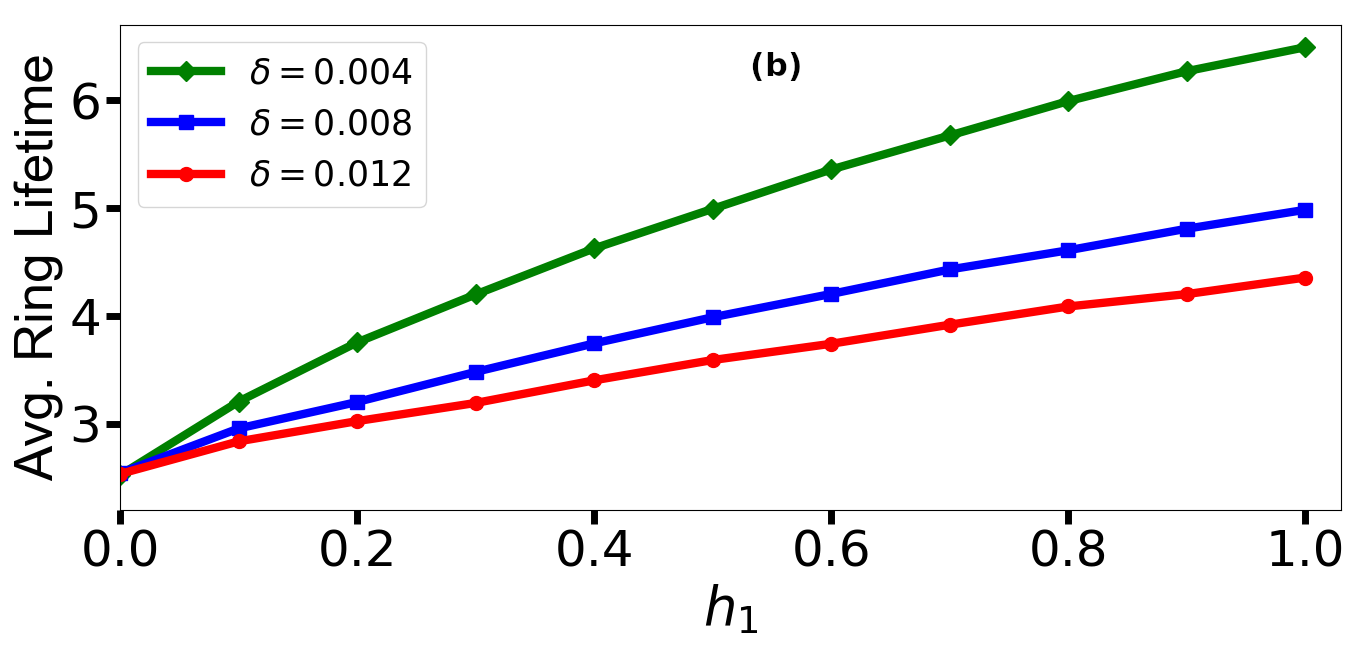}
\caption{The average ring-life time for single ring breaking are plotted as a function of hydrolysis rates. Data for polymerization rate 1.0 and for polymerization rate 4.0 are plotted in (a) and (b) respectively. All the plots are for random breaking and the different breaking rates $\delta$ are represented by different colors.}
\label{for2}
\end{figure}

\begin{figure}
\includegraphics[width=0.5\textwidth]{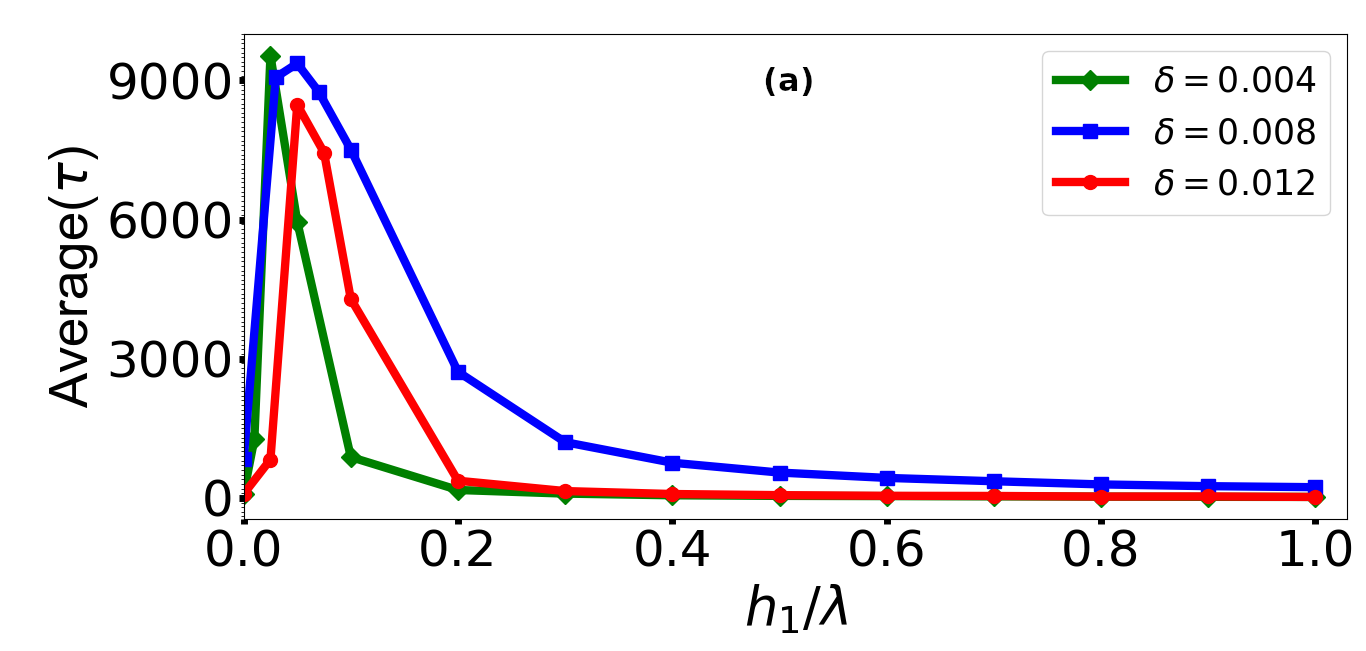}
\includegraphics[width=0.5\textwidth]{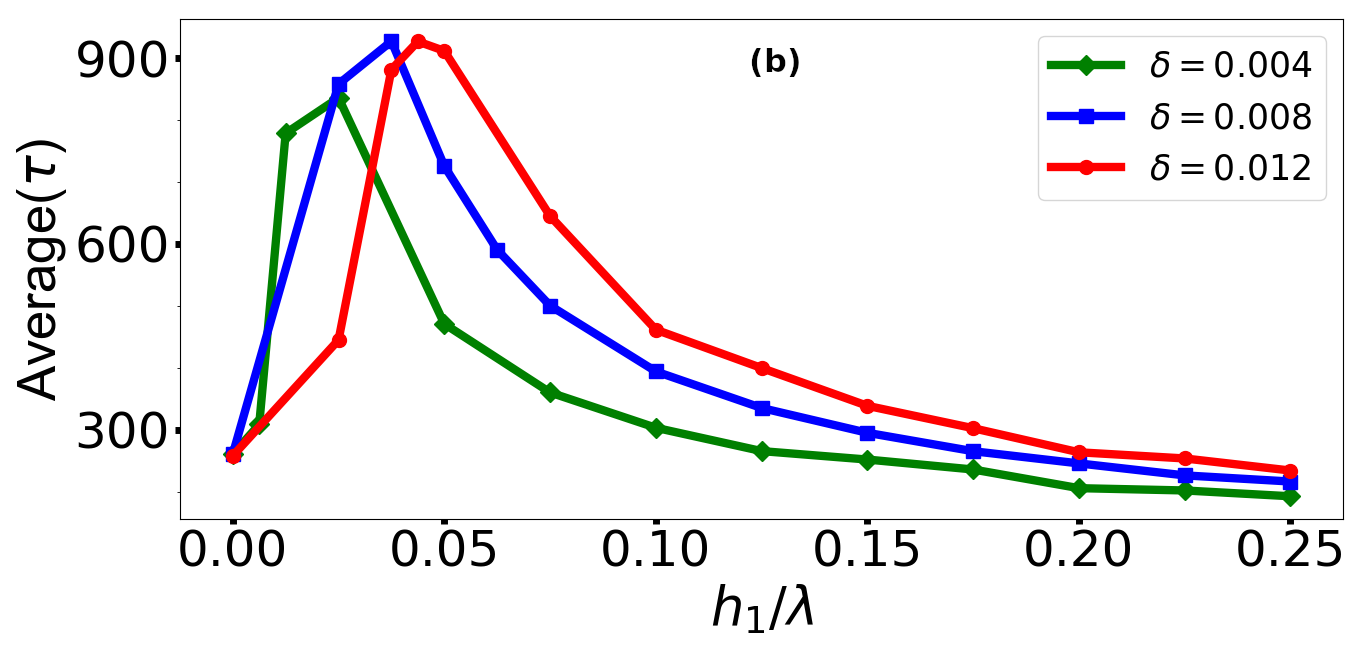}
\caption{The average lifetime ($\tau$) of FtsZ for irreversible ring opening are plotted as a function of hydrolysis rates.
Data for polymerization rate 1.0 and for polymerization rate 4.0 are plotted in (a) and (b) respectively. All the plots are for random breaking and the different breaking rates $\delta$ are represented by different colors.}
\label{for3}
\end{figure}

 We studied the effect of ring breaking rate on the FtsZ ring for random breaking 
model with vectorial hydrolysis. As the breaking rate increases, the average 
length of the ring for a given hydrolysis and polymerization rate increases(see Fig. \ref{for1}). Also as the breaking rate increases, the average length for single ring breaking of the polymer for a given hydrolysis and polymerization rate  decreases. For polymerization rate 1.0 different breaking rates i.e 0.008 and 0.012 
were found to be same. This suggests that for a given polymerization rate beyond a certain value, the breaking rate does not affect the average time for single ring breaking(see Fig. \ref{for2}). We also looked at the average time of irreversible opening. We find that as the breaking rate increases, the 
hydrolysis rate at which the average time for irreversible opening becomes maximum increases( Fig. \ref{for3}). 


\subsection{\label{sh}Stochastic hydrolysis(SH)}

SH in the case of microtubules provides a built-in mechanism for rescue and catastrophe and hence for the dynamics. This feature makes SH more likely than VH for microtubules. In this section we look at the FtsZ ring with SH. We will hence put $h_1=0$ now. 

\begin{figure}
\includegraphics[width=0.5\textwidth]{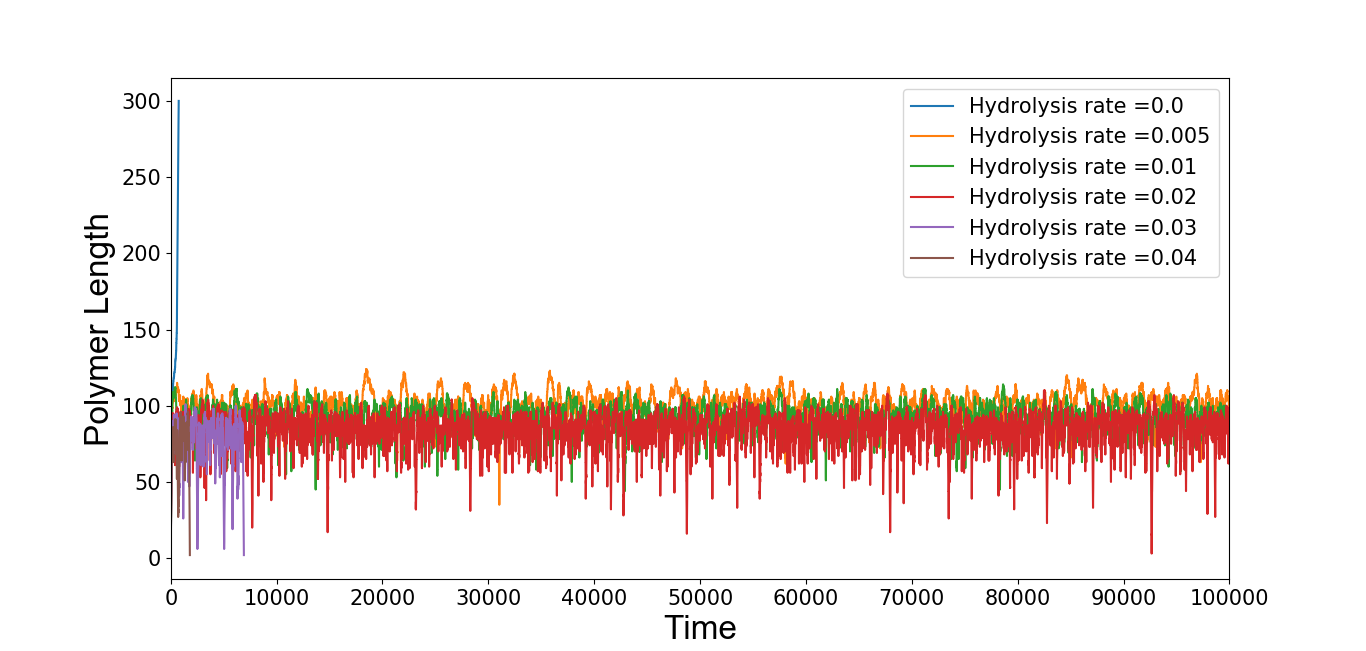}
\includegraphics[width=0.5\textwidth]{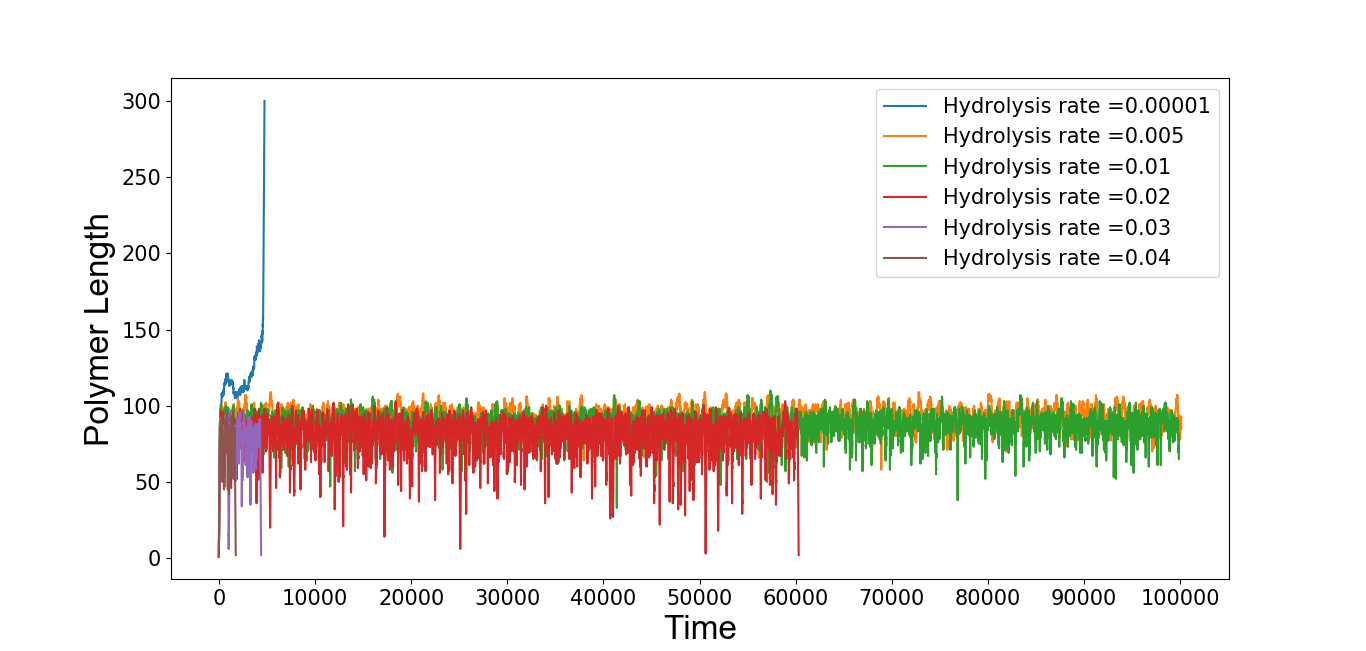}
\includegraphics[width=0.5\textwidth]{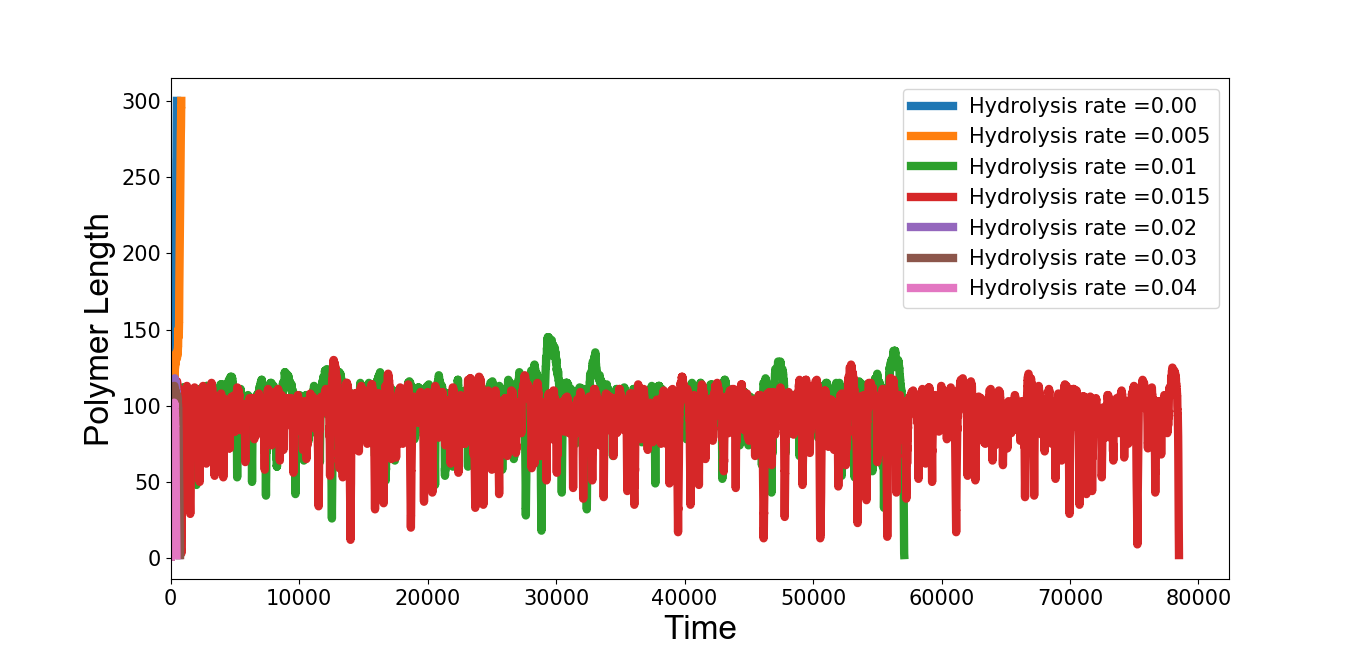}
\includegraphics[width=0.5\textwidth]{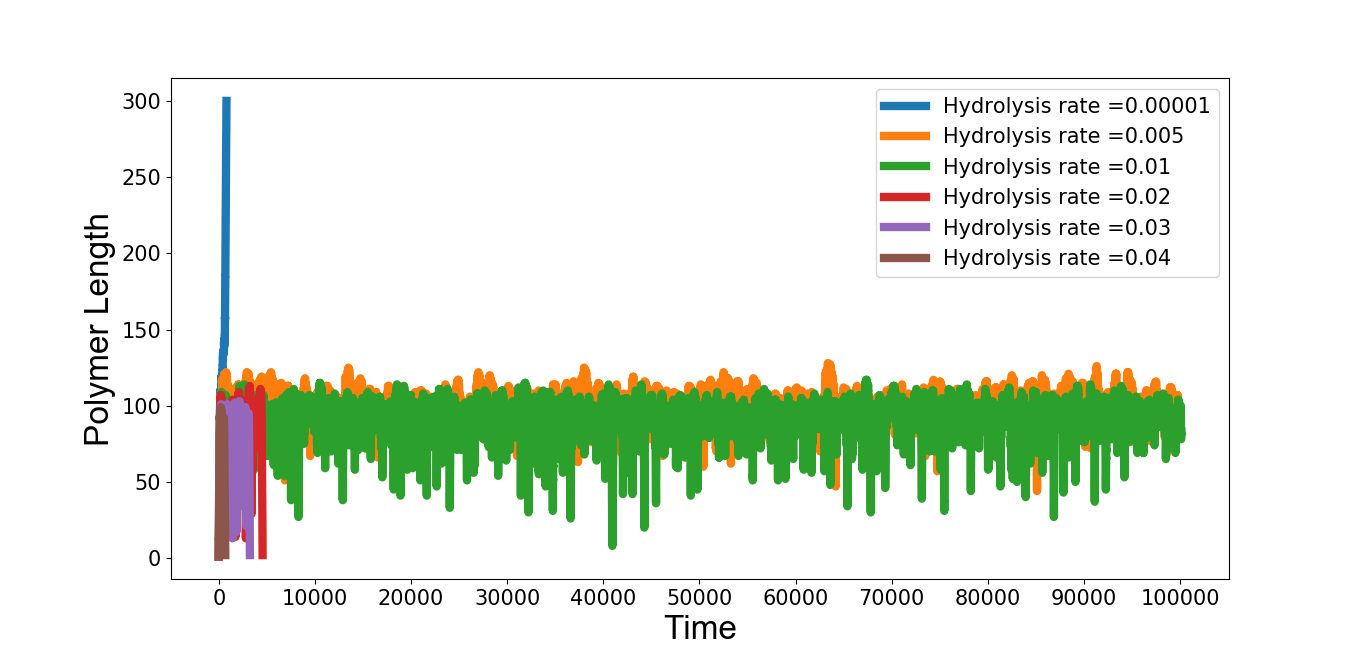}
\includegraphics[width=0.5\textwidth]{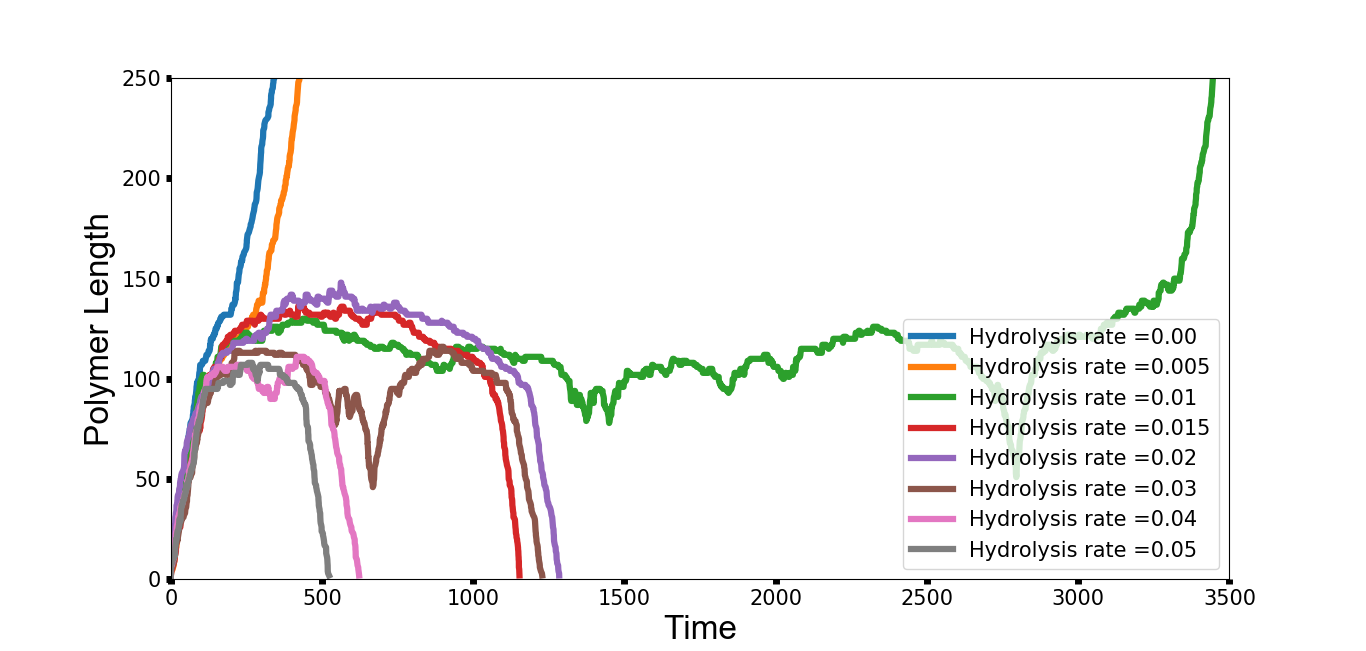}
\includegraphics[width=0.5\textwidth]{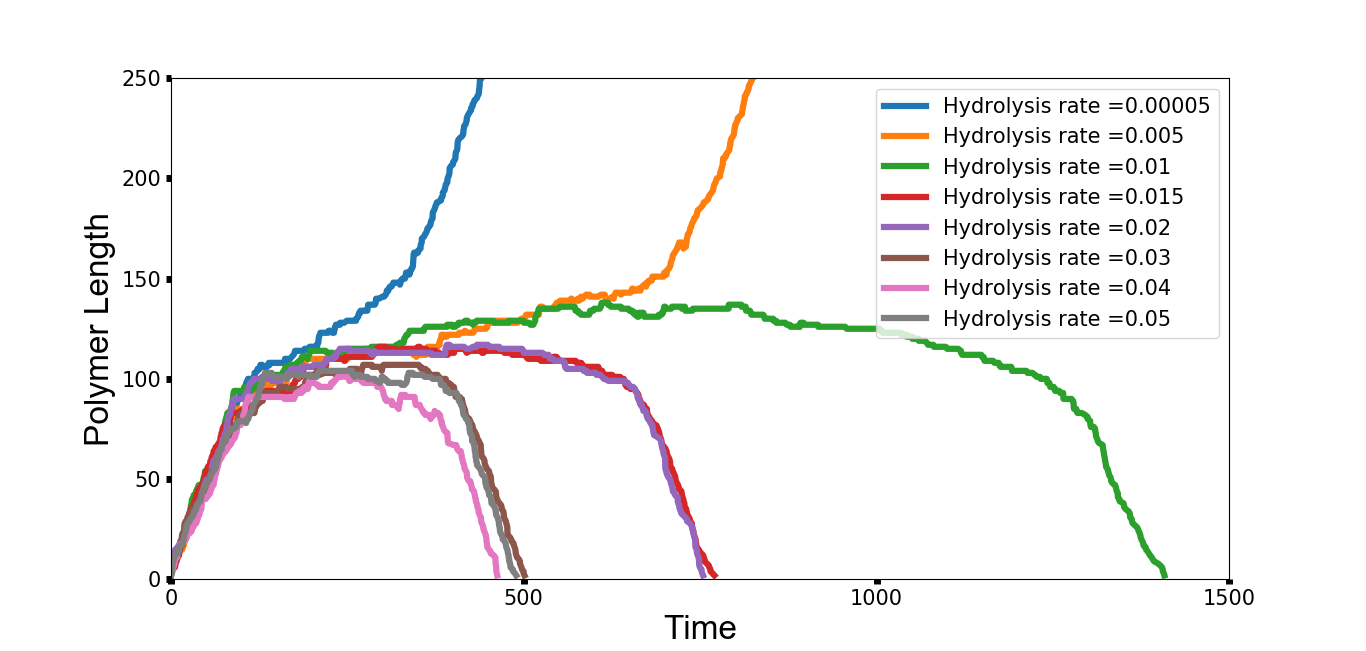}
\caption{Trajectories for random (left) and non-random breaking (right)  for 
polymerization rate 2.0, 4.0 and 8.0 respectively (top to bottom)}
\label{strajectory}
\end{figure}

We have seen in the previous section, the ring breaking naturally provides a way to build rescue and catastrophe in the FtsZ ring. For the sake of completion, we explore FtsZ dynamics assuming SH in this section. Fig \ref{strajectory}  shows trajectory with both  random breaking and non-random breaking. In the case of random breaking, 
there are now two ways to introduce interfaces in the system. As expected, the ring is more dynamic in this case and at the critical value of the hydrolysis rate $h_2$, the ring stays stable for a very long time with regular opening and closing of 
the ring. In the case of non-random breaking the ring is similar, but the ring is not as stable as in the case of 
SH with random breaking.

\begin{figure}
\includegraphics[width=0.5\textwidth, height=0.15\textheight]{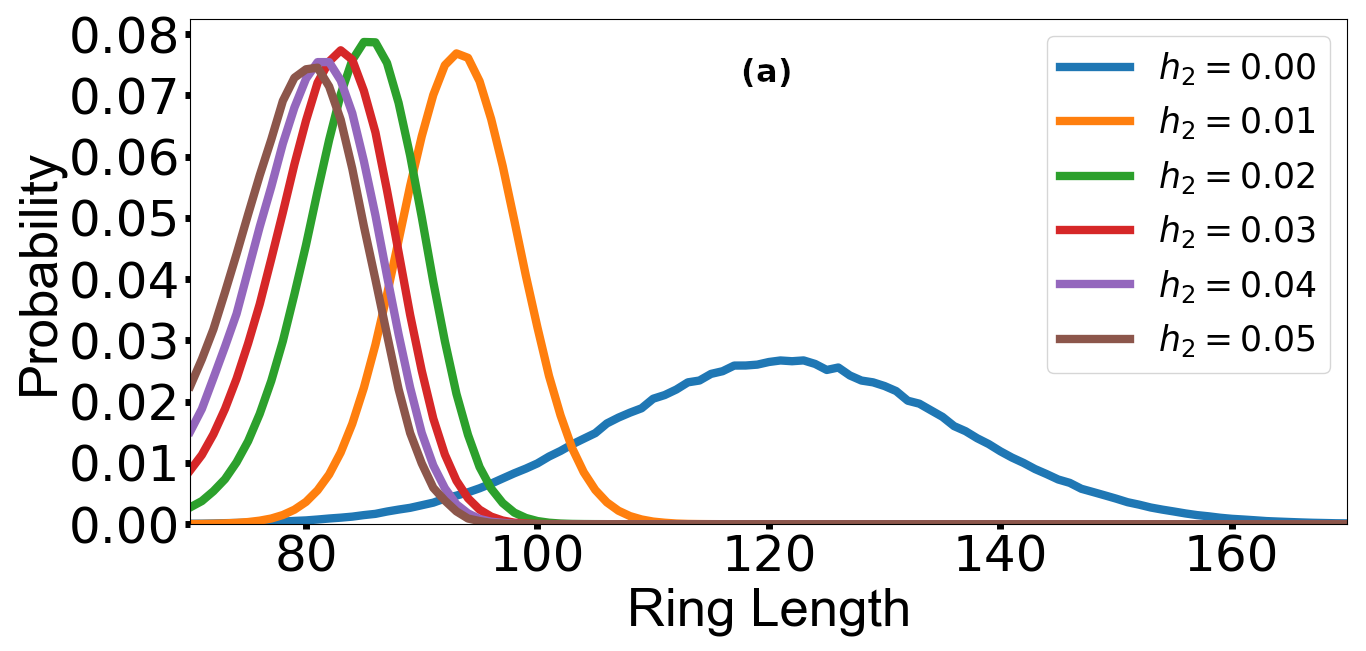}
\includegraphics[width=0.5\textwidth,height=0.15\textheight]{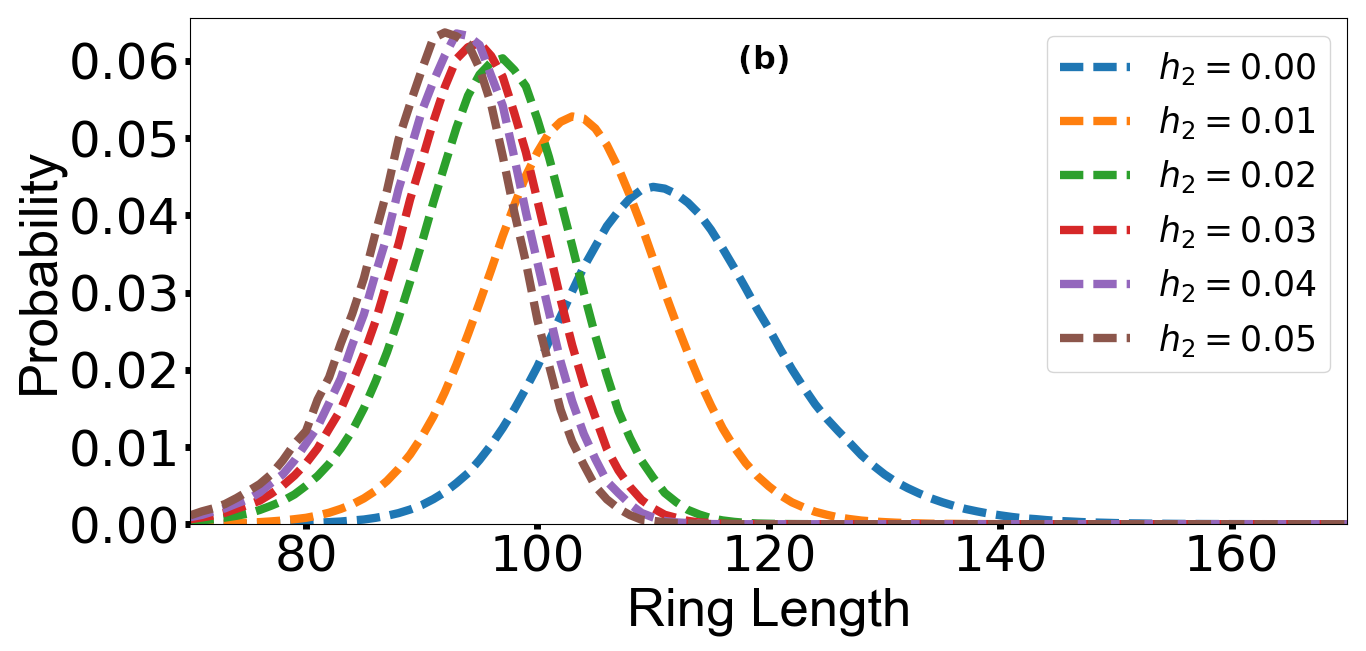}
\caption{Distribution of ring length for polymerization rate 4.0 for random(a) and non-random (b) breaking with stochastic hydrolysis. }
\label{slength}
\end{figure}

We looked at the ring length distribution in this case. We find that the ring length distribution is sensitive to change of hydrolysis rate for both , random and non-random breaking(see Fig. \ref{slength}).  However the non-random breaking case is less sensitive to the hydrolysis rate(especially for smaller polymerization rate). 

\begin{figure}
\includegraphics[width=0.5\textwidth]{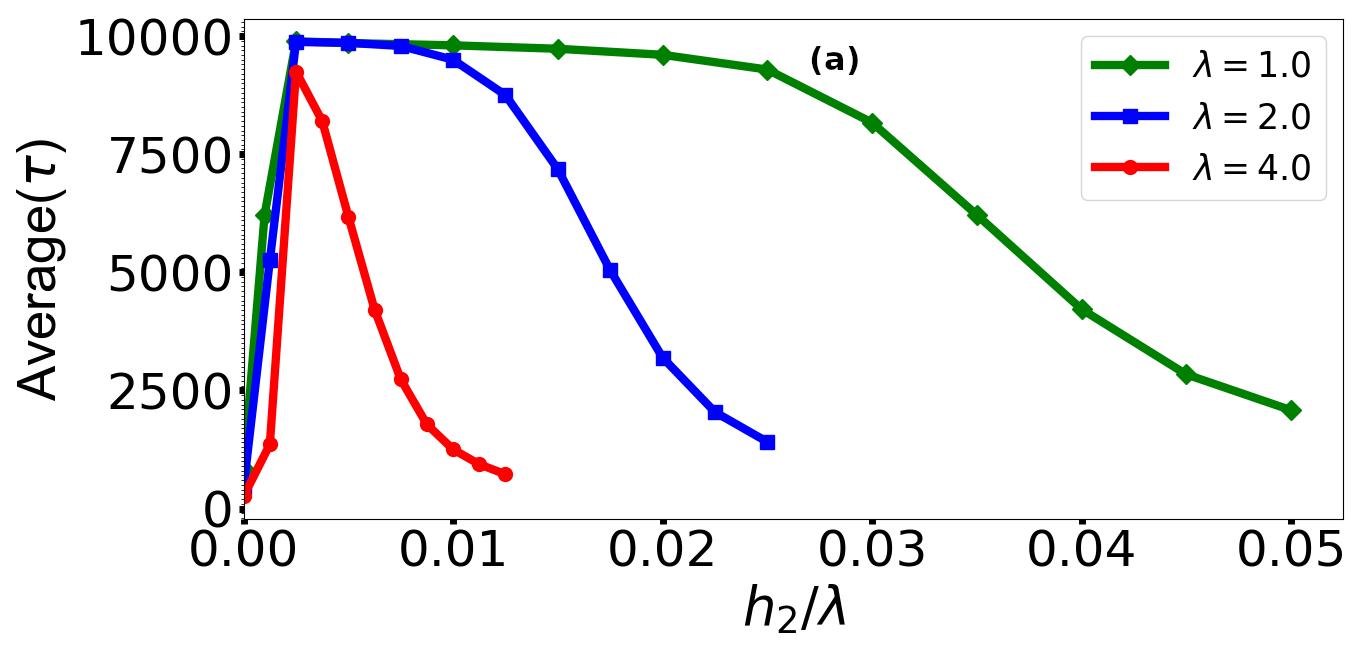}
\includegraphics[width=0.5\textwidth]{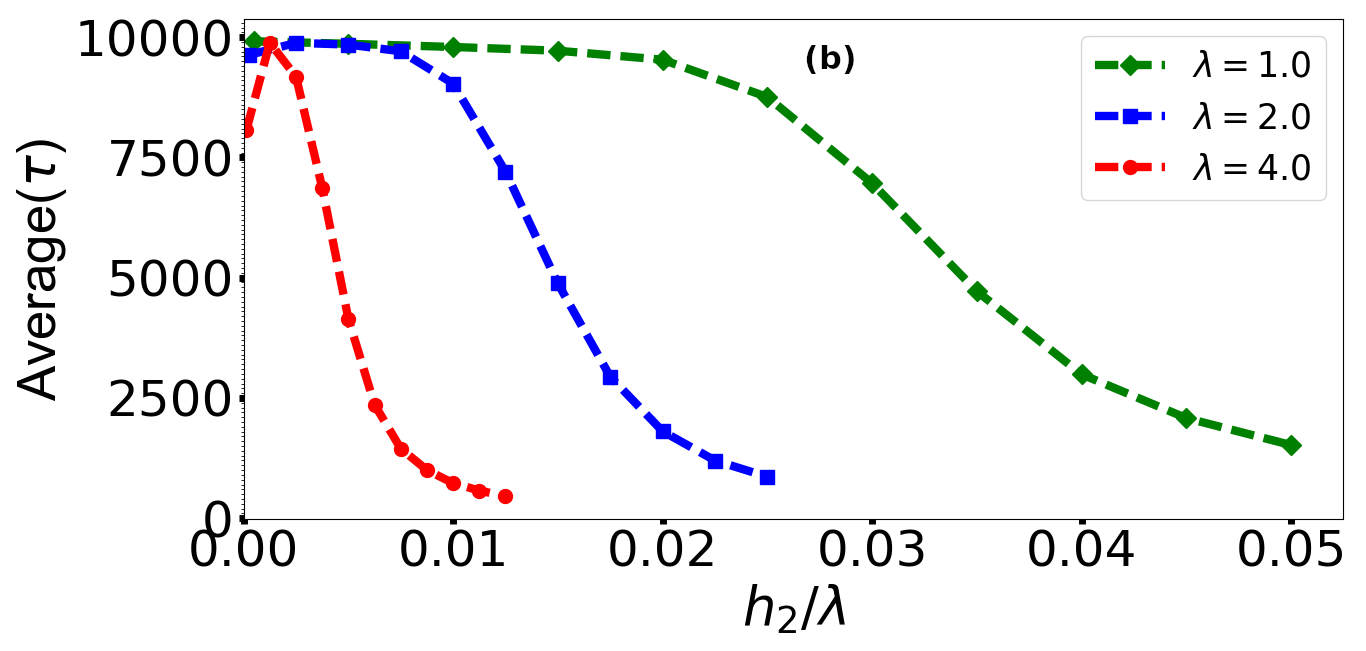}
\caption{Average lifetime of FtsZ rings before irreversible opening for random (a) and non-random (b) breaking with stochastic hydrolysis.}
\label{stime}
\end{figure}

We also looked at the time for irreversible ring opening(Fig. \ref{stime}) and interestingly we find that unlike the case of vectorial hydrolysis, the average 
lifetime is insensitive to the hydrolysis rate for a wide range of hydrolysis rate values.

\section{\label{concl}Conclusion}

The dynamics of FtsZ ring is being very actively studied in experiments in recent times\cite{tread1,tread2,Tread3,Tread4}. 
It is still not possible to observe the microscopic dynamics and one can at best study the average quantities like the average lifetime and length of the ring \cite{afm3}. In this paper, we have introduced a stochastic model which considers all the essential 
processes and compares all possible scenarios. This is important as the nature of hydrolysis and ring breaking is still not well understood, even though there is a general consensus that hydrolysis and treadmilling of the FtsZ polymer plays a crucial role in defining its dynamics and functionality. Note that the 
dynamic nature of the ring cannot be captured effectively in a deterministic model. Our stochastic model naturally gives rise to catastrophe and rescue of FtsZ ring, which have been observed in experiments. We find that unlike microtubules and actin the 
nature of hydrolysis does not play a crucial role for FtsZ ring. For microtubules and actin  if one considers vectorial hydrolysis , then one has to introduce rescues by hand for the polymer to be dynamic. We find that the rescues and catastrophe events are built 
in our model due to the ring topology irrespective of the hydrolysis mechanism. This suggests that ring breaking is a 
natural mechanism for introducing dynamics in the FtsZ filaments. 

In the case of vectorial hydrolysis with random ring breaking we could qualitatively reproduced 
the known behavior of the $Z$-ring. We found that increasing the hydrolysis rate in this case 
narrows the ring length distribution and the average length of the ring decreases with hydrolysis. 
This is consistent with the {\it{in vitro}} studies of Z-ring by Mateos-Gil {\it{et al.}} \cite{afm3}. We also find that the 
time lapsed between ring formation and final decay(time for irreversible ring opening) 
changes non-trivially with the hydrolysis rate. One needs to fine tune hydrolysis rate to achieve 
the most stable ring in this case. In contrast, when we consider non-random ring breaking, the behavior of the ring is insensitive to the hydrolysis rate.  Initially the polymer has only one interface where the hydrolysis occurs. But after ring closure, we find that in the case of random ring breaking, many new 
interfaces show up, and hence the ring can potentially break up at multiple places, consistent with the recent experiments \cite{Tread4}. Hence even though we consider a closed ring, in the case of random breaking, the ring actually has a patchy structure. In contrast in the case of non-random ring breaking, the  number of interfaces do not change with time and one typically has only one interface in the $Z$-ring.  All these observations  suggests that $Z$-ring breaks randomly. Similar conclusion was also reached by Mateos-Gil {\it et al.}  \cite{afm3}.  Recent works suggest that treadmilling of $Z$-ring is sufficient to generate the constriction force generated by $Z$-ring during cell division. The ring with random breaking is also contractile, as shown in Section \ref{ringbreak}. 

We have tried to keep our model basic so that we could distinguish the effect of different random processes clearly.  The model can be studied by including many other processes like allowing the attachment /detachment of short polymer and not just a single monomer during polymerization/ depolymerization process. We find that this does not change the picture qualitatively and hence not needed to understand the basic dynamics of the FtsZ polymer. One can also introduce lateral interaction to model the situation when  the FtsZ concentration is high, for example to model the dynamics ${\it in vivo}$.

We have also looked at the ring dynamics with stochastic hydrolysis. In this case we find that for both ring breaking mechanisms (random and non-random), we do not need to 
fine tune the hydrolysis rate to achieve the stable dynamic 
ring. Recent experiments \cite{invitro18} suggest that fine 
tuning of hydrolysis rate is required for stable dynamic ring. Behaviour of ring life time as a function of hydrolysis rate can hence be a distinguishing feature between vectorial and 
stochastic hydrolysis. Hence, differences in the statistics 
of the ring can be used  to unravel the actual mechanism of hydrolysis in the FtsZ polymer. 

We find that the randomness makes the  ring more dynamic and stable, as seen in the 
case of  stochastic hydrolysis and random breaking. In general it is now established that the stochasticity increases the stability of biological processes \cite{random} and is 
important to understand the functioning of biological systems. Though the Z-ring 
has been studied by others using deterministic models, this is the first stochastic model for 
Z-ring. We find that the model is able to explain a number of features of FtsZ filament dynamics {\it in vitro}, elucidating the crucial role the ring topology plays in their functioning. 
The model successfully brings out the role of hydrolysis and ring breaking mechanisms in forming a contractile 
dynamic ring. The model is simple and in future can easily be used for more quantitative studies or involve more 
complex interactions to study the {\it in vivo} dynamics of Z-ring. 

\section*{Acknowledgements}
 The authors thank Dr. R. Sreenivasan and Dr. S. Roychowdhury for fruitful discussions. This work has been 
 supported by Department of Atomic Energy, India thorugh the 12th plan project (12-R\&D-NIS-5.02-0100).

\section*{Author Contributions}
A.S ran the simulations. S and A.V.A.K developed the model. A.S,S and A.V.A.K analysed the results and drafted the manuscript.


\end{document}